\documentclass[11pt,letterpaper]{article}

\usepackage[T1]{fontenc}
\usepackage{lmodern}
\usepackage{microtype}

\usepackage[letterpaper,margin=1in]{geometry}
\usepackage{setspace}
\usepackage{authblk}

\usepackage{amsmath,amssymb,amsfonts}

\usepackage{graphicx}
\usepackage{booktabs}
\usepackage{array}
\usepackage{tabularx}
\newcolumntype{L}{>{\raggedright\arraybackslash}X}
\usepackage{float}
\usepackage{changepage}
\usepackage[table,dvipsnames]{xcolor}
\usepackage[aboveskip=1pt,labelfont=bf,labelsep=period,justification=raggedright,singlelinecheck=off]{caption}

\usepackage[ruled,vlined]{algorithm2e}

\usepackage{cite}
\usepackage{nameref}
\usepackage{hyperref}

\definecolor{CarveGeneralizability}{HTML}{167A6B}
\definecolor{CarveStability}{HTML}{C96E22}

\newcolumntype{+}{!{\vrule width 2pt}}

\newlength\savedwidth

\usepackage{authblk}

\newcommand{\carveY}{\checkmark}
\newcommand{\carveN}{\ensuremath{\times}}
\newcommand{\carveP}{\ensuremath{\triangle}}
\newcommand{\carveNA}{\textemdash}

\begin{document}

\title{Cluster Analysis with Resampling for Validation and Exploration (CARVE)}

\author[1,2]{Kai R. Wycik}
\author[4]{Tiffany M. Tang}
\author[5,$\dagger$]{Tarek M. Zikry}
\author[1,2,3,$\dagger$]{Genevera I. Allen}

\affil[1]{Department of Statistics, Columbia University, New York, NY, USA}
\affil[2]{Center for Theoretical Neuroscience, Zuckerman Mind Brain Behavior Institute, Columbia University, New York, NY, USA}
\affil[3]{Irving Institute for Cancer Dynamics, Columbia University, New York, NY, USA}
\affil[4]{Department of Applied and Computational Mathematics and Statistics, University of Notre Dame, Notre Dame, IN, USA}
\affil[5]{School of Data and Information Sciences, University of North Carolina at Chapel Hill, Chapel Hill, NC, USA}

\affil[$\dagger$]{Corresponding authors: \href{mailto:tarek@unc.edu}{\texttt{tarek@unc.edu}}; \href{mailto:genevera.allen@columbia.edu}{\texttt{genevera.allen@columbia.edu}}}

\date{}
\maketitle

\begin{abstract}
Clustering is widely used across the sciences as the foundation for downstream data-driven scientific discoveries. However, clustering results are highly sensitive to the choice of algorithm, preprocessing, and the number of clusters $k$, producing scientific claims that are often not reproducible. The current state of the art for validating clustering solutions consists of clustering validation indices (CVIs) such as Silhouette, Davies--Bouldin, and Calinski--Harabasz, which rely on geometric assumptions that break down on the heavy-tailed, high-dimensional, and nonlinearly structured data encountered in biomedical research. Resampling-based alternatives -- grounded in the ideas of clustering stability and generalizability -- have been proposed but remain scattered across specialized tools with no unified, accessible software. We fill this gap with CARVE (Cluster Analysis with Resampling for Validation and Exploration), an open-source Python and R package\renewcommand{\thefootnote}{\fnsymbol{footnote}}\footnote{\url{https://github.com/DataSlingers/CARVE/}}\renewcommand{\thefootnote}{\arabic{footnote}} that jointly evaluates multiple clustering algorithms and hyperparameters, returning stability and generalizability diagnostics at the global, cluster, and sample level together with principled selection rules and consensus-based cluster labels. Across six synthetic benchmarks CARVE consistently recovers near-optimal clusterings where classical indices degrade substantially. On experimental genomics and proteomics data sets, CARVE recovers finer biological structure when classical CVIs collapse entirely. CARVE is available with a scikit-learn-compatible Python API and an analogous R interface compatible with Seurat workflows.

\end{abstract}

\section{Introduction}
Clustering is a widely used class of unsupervised learning methods that partitions data into groups of similar samples. It is commonly used to derive data-driven discoveries in a wide variety of scientific fields \cite{allen2023interpretable}, including genetics \cite{do2008clustering, oyelade2016clustering, bikku2023optimizing}, neuroscience and psychiatry \cite{wiecki2015model, sun2015two, alashwal2019application}, social sciences \cite{zhang2024image, alsayat2016social, grimmer2021machine}, economics and environmental sciences \cite{wielechowski2021interdependence, noviandy2024environmental, vinska2021cluster}, astronomy \cite{fraix2021unsupervised, hawkins2015using, yu2022hierarchical}, and physics \cite{ali2021integration, mikuni2021unsupervised, francisco2026physics}. For example, clustering can be used to identify novel cell types in single-cell RNA-seq data \cite{do2008clustering}, characterize clinically significant conditions or subgroups of individuals \cite{wiecki2015model}, or uncover novel latent thematic groups or communities in textual and behavioral data \cite{grimmer2021machine, alsayat2016social}.

\begin{figure}[!h]
    \centering
    \includegraphics[width=1\linewidth]{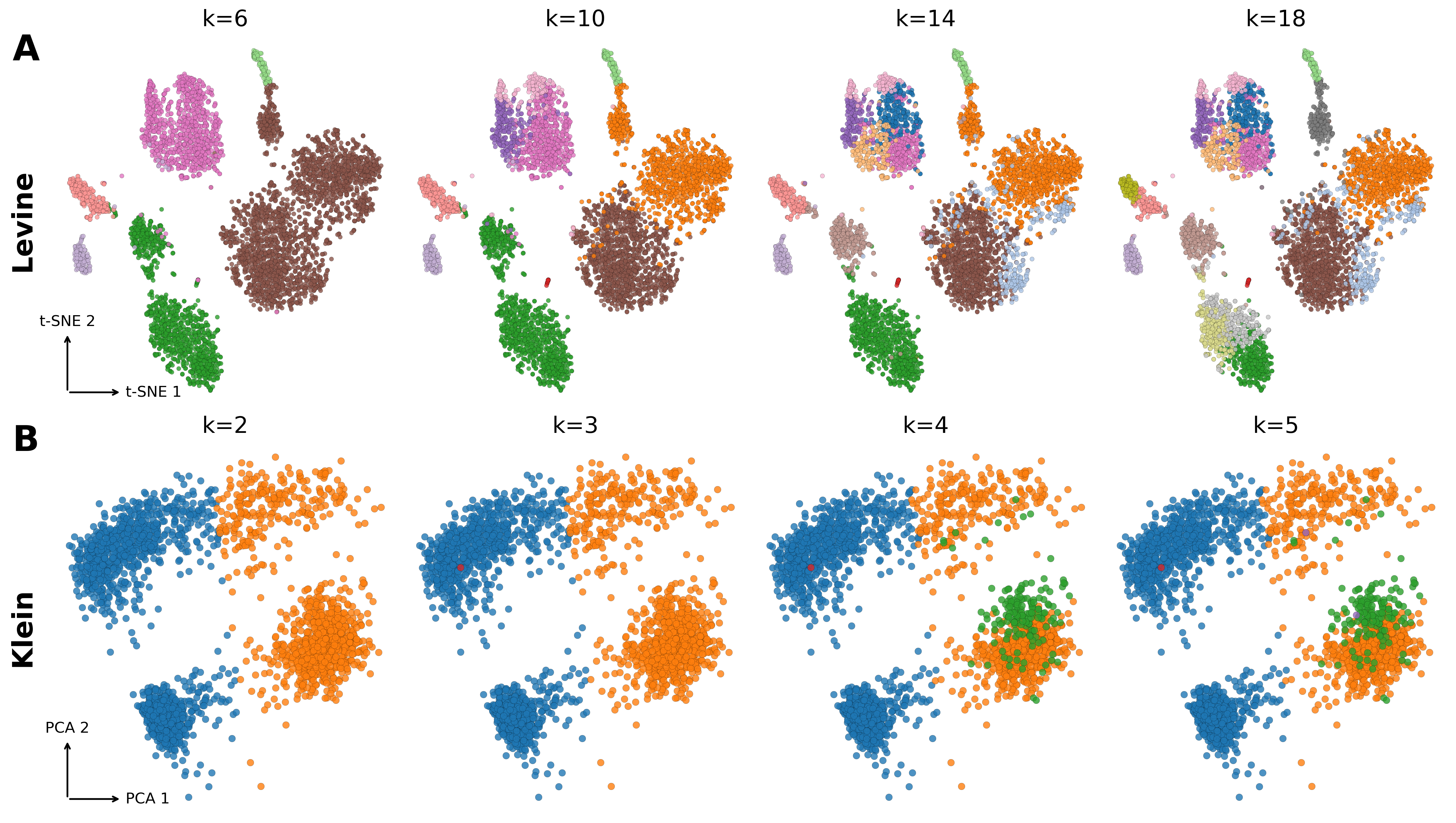}
    \caption{\textbf{Dependence of clustering results on the number of clusters $k$.}
    Two-dimensional embeddings colored by $k$-means \cite{lloyd1982least} cluster assignments for varying $k$. \textbf{(A)} t-SNE \cite{vandermaaten2008visualizing} embedding of the Levine 32dim mass cytometry dataset \cite{levine2015data} (top row, KMeans \cite{lloyd1982least} with $k \in \{6, 10, 14, 18\}$). \textbf{(B)} PCA \cite{hotelling1933analysis} embedding of the Klein droplet-based scRNA-seq dataset \cite{klein2015droplet} (bottom row, KMeans with $k \in \{2, 3, 4, 5\}$).
    In a biological context, each of these solutions would imply a qualitatively different number of cell types or cell stages, leading to fundamentally different scientific conclusions drawn from the same data.}
    \label{fig:clustering_ambiguity}
\end{figure}

In practice however, researchers often face a myriad of methodological choices that each lead to different clustering results, raising concerns about which results to trust. A particularly consequential choice is the number of clusters $k$. \textbf{Fig~\ref{fig:clustering_ambiguity}} illustrates this sensitivity on two datasets. Here, holding the clustering algorithm ($k$-means \cite{lloyd1982least}) and preprocessing pipeline fixed within each panel and varying only $k$ produces qualitatively different groupings. In a biological context, these differences imply fundamentally different numbers of cell types or developmental stages and lead to substantively different scientific conclusions \cite{gan2025machine, chang2025unsupervised, gibson2022perspectives, freytag2017cluster, crow2019single, chari2023specious}. 

To make matters worse, this sensitivity issue is exacerbated by other choices including the choice of algorithm, preprocessing steps, and additional hyperparameters. Clustering algorithms span many families---partitional methods such as $k$-means \cite{lloyd1982least} and $k$-medoids \cite{kaufman2009finding}, hierarchical agglomerative methods with different linkage criteria \cite{ward1963hierarchical}, spectral methods that operate on graph affinities \cite{von2007tutorial}, density-based methods such as DBSCAN \cite{ester1996density} and HDBSCAN \cite{campello2013density}, model-based methods such as Gaussian mixture models \cite{bishop2006pattern}, and graph-community methods such as Leiden \cite{traag2019louvain}---each bringing its own set of hyper-parameters beyond $k$ (e.g., linkage type, bandwidth, number of nearest neighbors, or resolution parameter). Moreover, before clustering even begins, choices about normalization, feature selection, and dimensionality reduction further expand the space of possible analysis pipelines. For complex biological data, which often exhibit non-spherical cluster geometries, heavy-tailed distributions, and high dimensionality \cite{chari2023specious, crow2019single}, there is no consensus on which combination of method and preprocessing is appropriate; instead, practitioners rely on field-specific conventions and rules of thumb that are rarely validated systematically \cite{gibson2022perspectives, freytag2017cluster}.

Given this large and opaque space of decisions, there is a need to evaluate the quality and reliability of clustering solutions in a principled way. We refer to this process as \textit{validation}, borrowing the term from its long-standing usage in the empirical sciences, where an experimental finding is considered trustworthy only after it has been independently checked, replicated, and shown to behave consistently under perturbations of the underlying conditions. Translating this notion to machine learning, validation assesses the replicability and reliability of results produced by a learning procedure \cite{allen2023interpretable}. Unlike supervised learning, however, clustering has no ground truth, so practitioners must rely on surrogate criteria to validate whether a solution reflects genuine structure or is merely an artifact of a particular methodological choice \cite{chang2025unsupervised}. 

The most widely used of such surrogate criteria are clustering validation indices (CVIs) \cite{rousseeuw1987silhouettes, davies2009cluster, calinski1974dendrite, tibshirani2001estimating}, which summarize a clustering with a single scalar score vis-\`a-vis geometric measures of compactness and separation (see \nameref{sec:related_work}). However, since these indices rely on strong distributional and geometric assumptions, which are usually violated in real world applications \cite{von2012clustering}, their ability to identify suitable clustering solutions on biological data sets is poor. There are multiple lines of work showcasing systemic over- or underclustering of CVIs on transcriptomic data \cite{dudoit2002prediction}, scRNA-seq benchmarks \cite{fang2022clustering, rautenstrauch2025shortcomings}, or on overlapping instead of well-separated clusters \cite{arbelaitz2013extensive, gagolewski2021cluster, vendramin2010relative, ikotun2025benchmarking}.

By contrast, resampling-based approaches aim to evaluate clusters without making parametric assumptions by quantifying whether the clusters persist across different data subsamples or perturbations. Two key resampling-based validation metrics are \textbf{stability} (whether the same groupings are discoverable when the data are subsampled, bootstrapped, or otherwise perturbed \cite{monti2003consensus, ben2001stability, dudoit2003bagging, kerr2001bootstrapping, levine2001resampling, fang2012selection, jain1987bootstrap, steinley2008stability, yu2019bootstrapping, li2019optimal, hennig2007cluster, liu2022stability}) and \textbf{generalizability} (sometimes called ``predictability''; whether cluster labels learned on one subset of the data can be accurately predicted on held-out samples \cite{dudoit2002prediction, lange2004stability, tibshirani2005cluster, tarekegn2025new, ullmann2022validation}). Stability and generalizability metrics are model-agnostic and do not rely on underlying assumptions with regards to cluster geometry, distribution, or clustering method used. Further, stability is a necessary condition for data-driven scientific discoveries, in particular, clustering interpretations to be considered reliable and trustworthy \cite{gan2025machine}. However, stability in itself is not sufficient, as simple partitions may be stable without providing meaningful scientific insight. Thus, generalizability is necessary to verify whether the insights derived from a subsample of the data generalize to held-out data. 

There is also a need to evaluate clustering solutions at different levels of granularity. At the global level, one assesses the overall quality of the cluster memberships across all observations. At the cluster and sample level, one identifies which individual clusters and samples are robust and which may be dubious. This is an important diagnostic given that the stability or generalizability of clusters and samples can vary substantially even within the same clustering solution \cite{hennig2007cluster}. Together, stability, generalizability, and multi-level analyses provide necessary perspectives for practitioners of cluster analyses to derive meaningful scientific insights from their data. 

\begin{figure}[!h]
    \centering
    \includegraphics[width=1\linewidth]{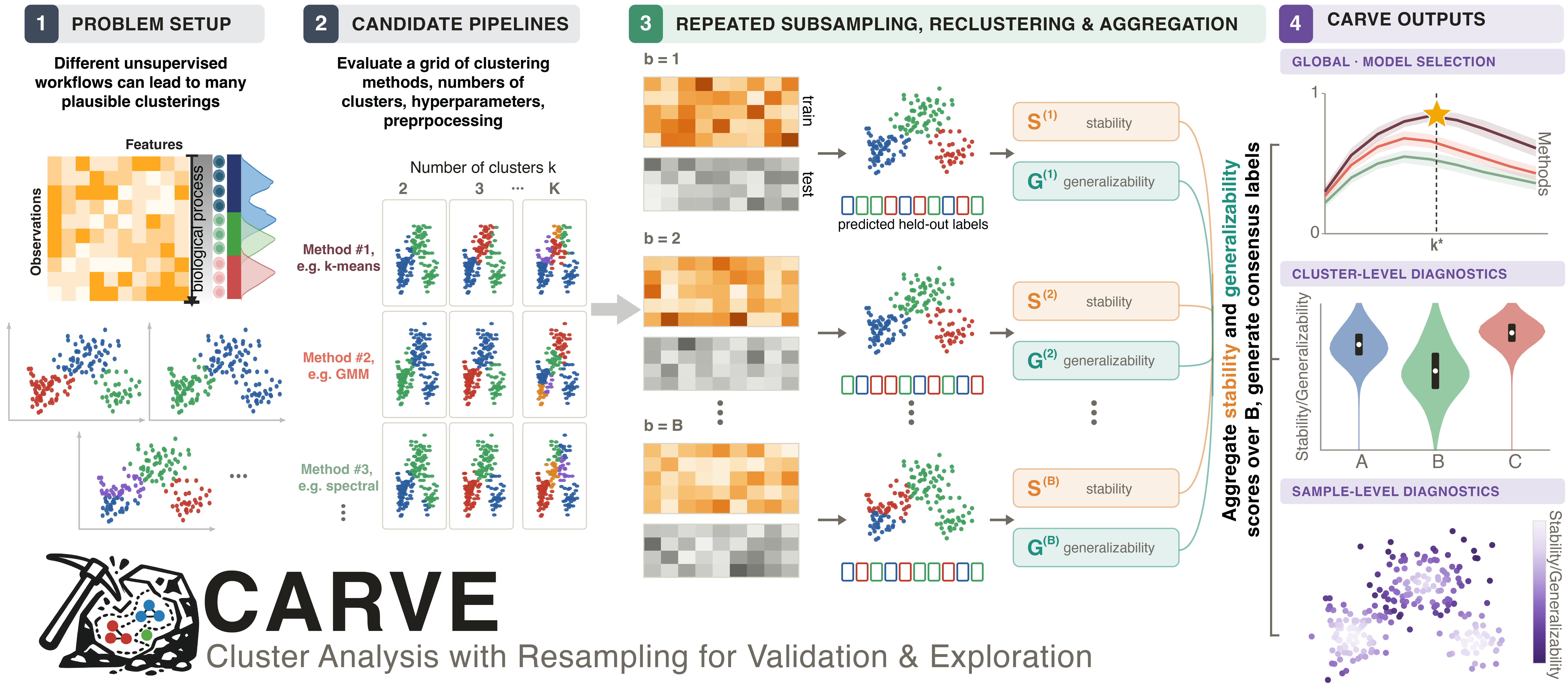}
    \caption{\textbf{Overview of CARVE.} Biological processes often form continua and admit many plausible clustering solutions from different workflows. CARVE takes as input a sample-by-feature data matrix together with a user-supplied grid of candidate clustering pipelines. From repeated resamples and reclusterings, CARVE computes two scores on global, cluster, and sample levels: stability (robustness of clustering solution to subsampling) and generalizability (predictability of clustering solutions to held-out data). These scores are aggregated to produce global validation curves over the number of clusters $k$ to support model and cluster-number selection and quantify cluster- and sample-level diagnostics to rigorously elucidate scientific insights. }
    \label{fig:carve_overview}
\end{figure}

For these reasons, we introduce CARVE (Cluster Analysis with Resampling for Validation and Exploration), a unified multi-level clustering validation framework that implements stability- and generalizability-based assessment across user-specified grids of clustering algorithms and hyperparameters. Importantly, CARVE produces diagnostics and visualizations at the global, cluster, and sample level, enabling systematic comparison and principled selection of both the number of clusters and the clustering method. We demonstrate CARVE on synthetic benchmarks and on real biomedical datasets, and provide open-source Python and R implementations to support integration into existing analysis pipelines (see \textbf{Fig~\ref{fig:carve_overview}} for an overview).

\section*{Related Work} \label{sec:related_work}
A wide range of clustering validation methods have been proposed in the literature (we refer the reader to comprehensive reviews for an overview of the full research area \cite{handl2005computational, karanikola2021investigating, liu2022stability, ullmann2022validation, hassan2024z, tomavsev2016clustering, ikotun2025cluster, halkidi2001clustering, todeschini2024validation, vendramin2010relative, von2010clustering, van2023white}). The most commonly used class of methods are clustering validation indices (CVIs), which assign a single scalar value to a clustering result that, when maximized or minimized, is intended to indicate a particularly suitable solution \cite{rousseeuw1987silhouettes, davies2009cluster, calinski1974dendrite, tibshirani2001estimating, hassan2024z, charrad2014nbclust}. However, many widely used CVIs, related validation methods, and their software implementations have significant shortcomings.

Prominent and widely used CVIs make significant and prohibitive statistical or geometrical assumptions about the distributions and shapes of clusters in the data, which are rarely met by data sets encountered in genomics and other biomedical fields. For example, the silhouette statistic \cite{rousseeuw1987silhouettes} assumes ratio-scale dissimilarities and prioritizes compact, well-separated, roughly spherical clusters. The Davies–Bouldin index \cite{davies2009cluster} assumes clusters have densities that decrease with distance from a centroid-like characteristic vector. The Calinski–Harabasz index \cite{calinski1974dendrite} is built around Euclidean sums-of-squares and explicitly targets minimum-variance spherical clusters. The Gap statistic \cite{tibshirani2001estimating} relies on comparing within-cluster dispersion to a reference null distribution. If these assumptions are not met by the data, as is often the case with real-world data sets, CVIs fail to identify suitable clustering solutions. In particular, \cite{dudoit2002prediction} show that the Gap statistic dramatically overestimates the number of clusters on cancer transcriptomic data. On the other hand, the Calinski--Harabasz and Davies--Bouldin indices tend to select too few clusters in scRNA-seq benchmarks \cite{fang2022clustering, rautenstrauch2025shortcomings}. Finally, large benchmarking studies have shown that CVIs' success rates drop dramatically when clusters are overlapping rather than well-separated, another common trait in real data \cite{arbelaitz2013extensive, gagolewski2021cluster, vendramin2010relative, ikotun2025benchmarking}.

Necessitated by the shortcomings of CVIs, resampling-based approaches evaluate clustering solutions by their robustness to perturbations. Stability criteria are based on the idea that meaningful structure should persist under subsampling, bootstrapping, or other data perturbations, and have been studied both at the global level and at finer granularity (e.g., cluster-wise or sample-level stability). This view underlies approaches such as consensus-style aggregation and stability via label agreement on overlapping subsamples among other variants \cite{monti2003consensus, ben2001stability, dudoit2003bagging, kerr2001bootstrapping, levine2001resampling, fang2012selection, jain1987bootstrap, steinley2008stability, yu2019bootstrapping, li2019optimal, hennig2007cluster, liu2022stability}. Another approach suggests to evaluate clustering solutions in terms of whether the partition generalizes: if cluster labels reflect reproducible structure in the data-generating process, then labels learned on one subset of the data should be transferable to held-out data, for example by training a classifier on training-set cluster labels and assessing predictive performance or label agreement on a test set \cite{dudoit2002prediction, lange2004stability, tibshirani2005cluster, tarekegn2025new, ullmann2022validation}. Both of these ideas align closely with the Predictability, Computability, and Stability (PCS) framework for veridical data science \cite{yu2020veridical}, which advocates that trustworthy data-driven findings should be both stable across data and modeling perturbations, predictive (or generalizable) on unseen observations, and computationally feasible for practitioners. Indeed, stability is a necessary condition for data-driven scientific discoveries. If clustering results are not stable under perturbations or subsampling, practitioners cannot meaningfully rely on or trust scientific claims made based on the respective cluster analysis. However, it often occurs that simple clustering solutions are disproportionately stable, while not providing meaningful scientific insight into the data. For this reason, it is important to jointly consider stability- and generalizability-based validation.

Single global validation scores are often insufficient for interpreting the data. For example, stability can vary substantially within the same fitted clustering, with some clusters appearing stable while others are dubious \cite{hennig2007cluster}. Related resampling approaches explicitly aim to quantify stability at finer granularity, including cluster- and sample-level stability \cite{yu2019bootstrapping}. Finally, while validating models on held-out or independently collected data is an established idea for supervised machine learning, it has historically received less systematic attention for clustering than CVIs, and existing methods often lack approachable software implementations \cite{allen2023interpretable, ullmann2022validation}.

The validation criteria described above remain largely siloed in practice. A researcher who wishes to jointly assess the stability and generalizability of a clustering solution, compare multiple algorithms and parameter settings, and diagnose quality at the global, cluster, and observation level must currently assemble an ad hoc pipeline from independently developed tools, each covering only part of the validation landscape \cite{brock2008clvalid, giancarlo2015valworkbench, landi2021reval}. This fragmentation makes it difficult to conduct validation in a standardized, reproducible way \cite{von2010clustering, hennig2015package}. The problem is compounded by the broader benchmarking culture in clustering: compared with supervised learning, the field lacks shared norms for method evaluation \cite{van2023white}, and published benchmarks frequently rely on only a handful of datasets \cite{gagolewski2022framework}, making it hard to draw general conclusions about when one validation strategy or clustering method should be preferred over another. 

\begin{table}[!h]
\begin{adjustwidth}{0in}{0in}
\centering
\scriptsize
\setlength{\tabcolsep}{3pt}
\renewcommand{\arraystretch}{1.15}

\begin{tabularx}{\textwidth}{@{} L c L ccc cccc @{}}
\toprule
& & &
\multicolumn{3}{c}{\textbf{Resolution}} & & & & \\
\cmidrule(lr){4-6}
\textbf{Package} &
\textbf{Lang.} &
\textbf{Approach} &
\textbf{G} & \textbf{C} & \textbf{S} &
\textbf{Stab.} & \textbf{Gen.} & \textbf{Multi} & \textbf{Plots} \\
\midrule

\multicolumn{10}{@{}l}{\textbf{Clustering validation tools}} \\
\midrule

\texttt{bootcluster} &
R &
Bootstrap stability \cite{yu2019bootstrapping} &
\carveY & \carveY & \carveY &
\carveY & \carveN & \carveN & \carveY \\

\texttt{fpc} &
R &
Bootstrap stability \cite{hennig2007cluster,tibshirani2005cluster} &
\carveY & \carveY & \carveN &
\carveY & \carveY & \carveN & \carveP \\

\texttt{ConsensusClusterPlus} &
R/Bioc &
Consensus clustering \cite{monti2003consensus} &
\carveY & \carveY & \carveY &
\carveY & \carveN & \carveN & \carveY \\

\texttt{consensusclustering} &
Py &
Consensus clustering \cite{monti2003consensus} &
\carveY & \carveN & \carveN &
\carveY & \carveN & \carveN & \carveY \\

\texttt{SC3} &
R/Bioc &
Consensus clustering \cite{kiselev2017sc3} &
\carveY & \carveN & \carveY &
\carveY & \carveN & \carveN & \carveY \\

\texttt{sharp} &
R &
Consensus clustering \cite{bodinier2023sharp} &
\carveY & \carveN & \carveN &
\carveY & \carveN & \carveN & \carveY \\

\texttt{M3C} &
R &
MC ref.\ consensus \cite{john2020m3c} &
\carveY & \carveN & \carveN &
\carveY & \carveN & \carveN & \carveY \\

\texttt{diceR} &
R &
Ensemble aggregation \cite{chiu2018dicer} &
\carveY & \carveN & \carveY &
\carveY & \carveN & \carveY & \carveY \\

\texttt{clue} &
R &
Ensemble / bagging \cite{hornik2005clue, dudoit2003bagging} &
\carveY & \carveN & \carveY &
\carveP & \carveN & \carveN & \carveY \\

\texttt{clusGap} &
R &
Gap statistic \cite{tibshirani2001estimating} &
\carveY & \carveN & \carveN &
\carveN & \carveN & \carveN & \carveY \\

\texttt{gapstatistics} &
Py &
Gap statistic \cite{tibshirani2001estimating} &
\carveY & \carveN & \carveN &
\carveN & \carveN & \carveN & \carveY \\

\texttt{clValid} &
R &
Index-based validation \cite{brock2008clvalid} &
\carveY & \carveN & \carveN &
\carveY & \carveN & \carveY & \carveY \\

\texttt{NbClust} &
R &
Index-based validation \cite{charrad2014nbclust} &
\carveY & \carveN & \carveN &
\carveN & \carveN & \carveN & \carveP \\

\texttt{pvclust} &
R &
Multiscale bootstrap \cite{suzuki2006pvclust} &
\carveN & \carveY & \carveN &
\carveY & \carveN & \carveN & \carveY \\

\texttt{OTclust} &
R &
OT mean partition \cite{li2019optimal,zhang2020cps} &
\carveY & \carveY & \carveY &
\carveP & \carveN & \carveN & \carveY \\

\texttt{Clest} &
R &
Prediction-based \cite{dudoit2002prediction} &
\carveY & \carveN & \carveN &
\carveN & \carveY & \carveN & \carveN \\

\texttt{prediction-strength} &
Py &
Prediction-based \cite{tibshirani2005cluster} &
\carveY & \carveN & \carveN &
\carveN & \carveY & \carveN & \carveN \\

\texttt{reval} &
Py &
Relative validation \cite{landi2021reval, lange2004stability} &
\carveY & \carveN & \carveN &
\carveY & \carveY & \carveY & \carveY \\

\texttt{clusterExperiment} &
R/Bioc &
Resampling ensemble \cite{risso2018clusterexperiment} &
\carveY & \carveY & \carveY &
\carveY & \carveN & \carveY & \carveY \\

\midrule
\multicolumn{10}{@{}l}{\textbf{Benchmarking \& index libraries}} \\
\midrule

\texttt{clustering-benchmarks} &
Py &
Benchmark suite \cite{gagolewski2022framework} &
\carveNA & \carveNA & \carveNA &
\carveNA & \carveNA & \carveY & \carveY \\

\texttt{ValWorkBench} &
Java &
Index library \cite{giancarlo2015valworkbench} &
\carveNA & \carveNA & \carveNA &
\carveNA & \carveNA & \carveN & \carveN \\

\midrule
\rowcolor{gray!10}
\textbf{CARVE} &
Py, R &
Stab.\ + gen.\ framework &
\carveY & \carveY & \carveY &
\carveY & \carveY & \carveY & \carveY \\
\bottomrule
\end{tabularx}

\caption{\textbf{Comparison of clustering validation software.}
G: global, C: cluster-level, S: sample-level.
Stab.: stability assessment; Gen.: generalizability assessment;
Multi: compares multiple clustering algorithms in one run; Plots: built-in visualization.
\carveY\ = yes; \carveN\ = no; \carveP\ = partial/limited; \carveNA\ = not applicable.}
\label{tab:software_cluster_validation_landscape}
\end{adjustwidth}
\end{table}

\subsection*{Contribution}
To address the fragmentation and limitations described above, we introduce CARVE (Cluster Analysis with Resampling for Validation and Exploration), a clustering validation package that rigorously evaluates clusterings over a user-specified grid of models and hyperparameters. Unlike widely used CVIs, CARVE relies on resampling-based stability and generalizability criteria that are model-agnostic and robust to cluster geometry or the underlying data distribution, making it well suited for complex, high-dimensional data commonly encountered in genomics and other biomedical fields. Rather than requiring analysts to assemble ad hoc pipelines from independently developed tools, CARVE unifies stability and generalizability assessment within a single framework, with built-in support for sweeping over multiple clustering algorithms and preprocessing configurations simultaneously (see Table~\ref{tab:software_cluster_validation_landscape}).

Concretely, CARVE produces (i) validation metric curves over $k$ for each candidate model, (ii) lightweight, transparent, yet principled selection of $k$ and other hyperparameters, (iii) robust cluster labels via consensus-based clustering, and (iv) diagnostics and visualizations that support interpretation at the global, cluster, and sample level. To further enhance reproducibility and transparency, CARVE standardizes the resampling procedure---including resampling, label-comparison metric, and pre-processing choices---and reports fully resolved configurations for each run. CARVE is available as an open-source Python library with a scikit-learn-style interface, as well as an R package, to support integration into existing analysis pipelines. We demonstrate CARVE on synthetic benchmarks and on real biomedical datasets, emphasizing settings where both the choice of $k$ and the choice of clustering algorithm affect scientific conclusions.

\section*{Design and implementation}

\subsection*{Methodology}
The goal of CARVE is to develop an integrated algorithm, evaluating clustering solutions with respect to stability and generalizability on multiple levels and to provide users with tools to explore and understand the results of the validation procedure. Our method is summarized in Algorithm \ref{alg:carve_stability_short}. 

CARVE takes as input a data matrix \(X \in \mathbb{R}^{n \times p}\) together with one or more candidate clustering configurations, each defined by a clustering estimator \(f\) and hyperparameters \(\theta\) (including the number of clusters \(k\)). Additional run-level parameters are the number of resampling iterations \(B\) and the subsampling proportion \(\rho \in (0,1)\). For each configuration, each iteration \(b\) draws two independent subsamples of size \(\lfloor \rho n \rfloor\), clusters each, and computes two main types of validation metrics: a per-iteration stability score \(S^{(b)}\), measured by the adjusted Rand index (ARI) \cite{hubert1985comparing} on the overlap of the two clusterings \cite{ben2001stability}, and a per-iteration generalizability score \(G^{(b)}\), computed by training a classifier model (a random forest \cite{breiman2001random} by default) on one subsample and evaluating it against a clustering of the held-out complement \cite{dudoit2002prediction}. Across iterations, CARVE aggregates \(\{S^{(b)}\}_{b=1}^B\) and \(\{G^{(b)}\}_{b=1}^B\) into means, standard errors, and 95\% confidence intervals; constructs a consensus matrix \cite{monti2003consensus} \(M \in [0,1]^{n \times n}\) from co-clustering frequencies (which further yields proportion of ambiguous clusters (PAC) \cite{senbabaouglu2014critical}, Gini, and cross-entropy stability summaries); and records a per-sample accuracy array \(E \in [0,1]^n\) whose mean is the global out-of-sample prediction accuracy. Further detail and pseudocode are given in \nameref{S1_Text}; details on user-facing inputs and outputs are provided in \nameref{S2_Text}.

\begin{algorithm}
\caption{CARVE computation for a given configuration $(f,\theta)$}
\label{alg:carve_stability_short}
\DontPrintSemicolon

\begingroup
\setstretch{1}
\footnotesize
\SetAlgoSkip{0.5em}
\SetAlgoNlRelativeSize{-1}

\KwIn{
Data matrix $X$; clustering method $f(\cdot, \, \theta)$ (including $k$);
resampling iterations $B$; subsampling proportion $\rho$;
}
\KwOut{
Stability scores $\{S_b\}_{b=1}^B$ (ARI stability);
consensus matrix $\widehat{M}_{f,\theta}$ and consensus-derived summaries
}

\BlankLine

\For{$b\gets 1$ \KwTo $B$}{
    \textbf{Sub-sample.} Draw two subsamples $P_1^{(b)}, P_2^{(b)}$, and let $P_\text{test}^{(b)} \gets X \setminus P_1^{(b)}$. \;
    \textbf{Cluster.} Use $f(\cdot, \, \theta)$ to obtain clusterings $C_1^{(b)}$, $C_2^{(b)}$, and $C_\text{test}^{(b)}$. \;

    \BlankLine \;

    \textcolor{CarveStability}{\textbf{Stability:}} \\
    \textcolor{CarveStability}{(i): \textbf{Global} $S^{(b)}$: ARI on overlap of $C_1^{(b)}$ and $C_2^{(b)}$.} \;
    \textcolor{CarveStability}{(ii): \textbf{Per-sample} update consensus matrix with $C_1^{(b)}$, $C_2^{(b)}$, aggregate row-wise after final update.} \; 

    \BlankLine \;

    \textcolor{CarveGeneralizability}{\textbf{Generalizability:} fit classifier $\hat{h}(\cdot)$ to $(P_1^{(b)}, C_1^{(b)})$, obtain $\hat{C}_\text{test}^{(b)} \gets \hat{h}(C_\text{test}^{(b)})$.} \\
    \textcolor{CarveGeneralizability}{(i): \textbf{Global} $G^{(b)}$: ARI of predicted $\hat{C}_\text{test}^{(b)}$ and held-out labels $C_\text{test}^{(b)}$.} \;
    \textcolor{CarveGeneralizability}{(ii): \textbf{Per-sample} record $C_\text{test}^{(b)}$, $\hat{C}_\text{test}^{(b)}$ to update per-sample generalizability.} \;
}

\BlankLine

\textbf{Aggregate.} Report the mean, standard errors, and $95\%$ quantiles of $\{S^{(b)}\}_{b=1}^B$ and $\{G^{(b)}\}_{b=1}^B$. Report consensus matrix, align clusterings $C_\text{test}^{(b)}$, $\hat{C}_\text{test}^{(b)}$ to report per-sample accuracy rates.

\endgroup
\end{algorithm}

\begin{figure}
    \centering
    \includegraphics[width=1\linewidth]{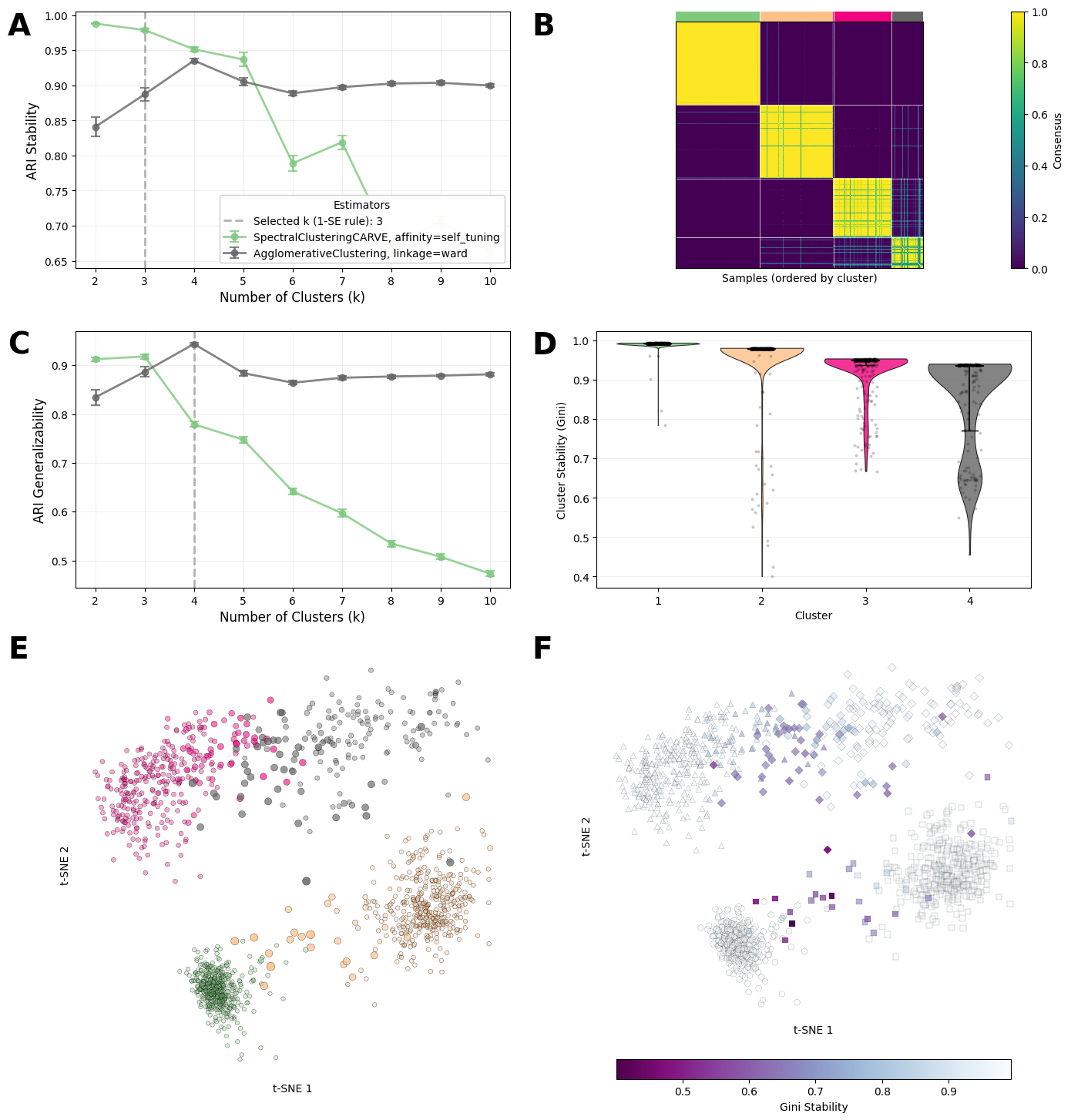}
    \caption{\textbf{Example CARVE output on the Klein droplet-based scRNA-seq dataset \cite{klein2015droplet}.}
    Visual output returned by a \texttt{CARVE.fit()} run that swept Ward agglomerative clustering and self-tuning spectral clustering across $k \in \{2, \dots, 10\}$. \textbf{(A)} Stability ARI as a function of $k$ for each estimator. The method selected by the respective selection-rule is marked by a vertical line. \textbf{(B)} Consensus matrix for a selected configuration: clear block-diagonal structure indicates samples that consistently co-cluster across resamples, while higher off-diagonal values indicate ambiguous samples and clusters. \textbf{(C)} Generalizability ARI as a function of $k$ for the same estimators, again with the 1-SE-selected model highlighted. \textbf{(D)} Per-cluster stability scores (violin plot) for the selected partition, exposing which of the ten clusters are stable across resamples and which are heterogeneous or dubious. \textbf{(E)} t-SNE \cite{vandermaaten2008visualizing} embedding colored by the CARVE-selected consensus labels. Dubious samples are larger with higher opacity. \textbf{(F)} t-SNE embedding marked by CARVE-selected consensus cluster-assignment. Highlighted samples are more spurious.}
    \label{fig:carve_output_klein}
\end{figure}

On a global level, stability and generalizability scores are assessed using the adjusted Rand index (ARI) \cite{hubert1985comparing}, a metric for comparing the agreement of two partitions of data. The ARI's domain is $[0, 1]$, with $0$ indicating no agreement between the partitions and $1$ indicating full agreement. Thus, we aim to maximize global stability, calculated as the sample mean $\frac{1}{B} \sum_{b=1}^{B} S^{(b)}$, and global generalizability, similarly calculated as $\frac{1}{B} \sum_{b=1}^{B} G^{(b)}$. However, inspired by how hyperparameters are commonly chosen in supervised learning \cite{hastie2009elements}, our benchmarking experiments show that choosing the largest $k$ within one standard error of the maximizing $k$ (1-SE rule) yields better results than merely maximizing the respective metric. In some settings, however, our benchmarks also show that the variation bands around the average ARI are so tight that the 1-SE rule becomes overly conservative and selects too few clusters. In such cases — for instance, on the Gaussian and $t$-distributed cluster simulations (see \nameref{S2_Table}, \nameref{S3_Table} and \nameref{S4_Table}) — a quantile-based rule performs better. Let $\bar{M}_k = \frac{1}{B} \sum_{b=1}^{B} M_k^{(b)}$ be the average of the respective metric $M$ (stability or generalizability) at $k$, and let $k^* = \arg\max_k \bar{M}_k$ be the maximizing configuration. We then select the largest $k$ whose average still lies within the central $95\%$ of the score distribution observed at $k^*$.

Co-clustering counts are aggregated into consensus matrices \cite{monti2003consensus} for each configuration. These are $n \times n$ matrices, in which each entry indicates the proportion of how often two samples were assigned the same cluster, divided by how often the two samples were sampled together. We reorder the rows of the consensus matrices via agglomerative clustering to reveal block-like structures. In very stable clustering solutions, one expects to see blocks along the diagonal of the matrix with values close to 1 with little to no weight in the off-diagonals. On stable samples, we expect row-wise entries to be mostly $0$ or $1$, while on unstable samples, we expect co-clustering proportions that are between $0$ and $1$. We thus obtain sample-level stability as the Gini index or cross-entropy of the respective row of the consensus matrix, belonging to that sample. For sample-level generalizability, after aligning labels, we may consider how often individual samples were misclassified by the classifier that predicts cluster labels on the held-out subsample. These sample-level stability and generalizability metrics are aggregated cluster-wise (by averaging) to obtain stability and generalizability metrics for the respective cluster.

Our benchmarking results indicate that CARVE's global stability metric with the 1-SE rule recovers clustering solutions with the highest agreement to ground truth cluster labels. Thus, we recommend using stability with the 1-SE rule as the default metric to select clustering solutions. Generalizability with the 1-SE rule should additionally be consulted if visualizations reveal that the clustering selected via stability with the 1-SE rule is overly simple. In practice, the choice of a good number of clusters should always be coupled to domain knowledge and incorporate both metrics in tandem. Consensus matrices, per-cluster stability and generalizability metrics, as well as per-sample stability and generalizability, should then be considered to decide between multiple potential clustering solutions, as well as to detect potentially dubious or continuous clusters which may require more careful interpretation.

\paragraph{Visual exploration.}
CARVE provides four visualization types for exploring validation results:

\begin{enumerate}
    \item \textbf{Metrics over \(k\) (global-level).}
    A summary plot showing CARVE's global stability or generalizability metric across values of $k$. Each line represents a combination of estimator and hyperparameter settings, with indicators showing $\pm 1$ standard error. The plot highlights the $k$ and configuration chosen by the max, 1-SE, or quantile rule with a dashed vertical line (see \textbf{Fig~\ref{fig:carve_output_klein}(A, C)}).

    \item \textbf{Consensus matrix heatmaps (cluster-level).}
    For the chosen clustering solution, CARVE plots the corresponding consensus matrix \cite{monti2003consensus} (after re-ordering samples to reveal block structure), providing a direct view of cluster stability and ambiguity in the resampling consensus (see \textbf{Fig~\ref{fig:carve_output_klein}(B)}). 

    \item \textbf{Box and violin plots (cluster-level).}
    For the chosen clustering solution, CARVE provides box- and violin plots of the respective metric (Gini (stability), CE (stability), and Accuracy (generalizability)) per cluster. This supports diagnosing which clusters are driving high or low validation scores  (see \textbf{Fig~\ref{fig:carve_output_klein}(D)}). 

    \item \textbf{Scatter plots (cluster and sample-level).}
    For the chosen clustering solution, CARVE also provides scatter plots which highlight samples which have uncertain cluster membership. Users may rely on standard embedding techniques (e.g., PCA \cite{hotelling1933analysis}, t-SNE \cite{vandermaaten2008visualizing}, UMAP \cite{mcinnes2018umap}) to visualize these uncertainties or supply their own proprietary embeddings. This supports diagnosing which samples are driving high or low validation scores (see \textbf{Fig~\ref{fig:carve_output_klein}(E, F)}).
\end{enumerate}

\section*{Results}

\subsection*{Overview of the evaluation design}
We evaluated CARVE by conducting simulation benchmarks where the true labels and true number of clusters $k^\star$ are known, allowing us to quantify how well each selection criterion recovers the underlying partition and by presenting real-data case studies that illustrate how CARVE supports practical model selection and interpretation beyond a single global performance score.

\subsection*{Synthetic benchmarking}
We benchmarked CARVE across six clustering tasks: isotropic Gaussian mixtures, heavy-tailed $t$-mixtures, $t$-mixtures with up to 1{,}536 nuisance features, and three nonlinearly embedded manifolds (swiss rolls, circles, moons) mapped to the observed space via random Fourier features (RFF) each with $k^\star = 5$ clusters. We use an appropriate clustering method for each setting: KMeans for Gaussians, Ward agglomerative clustering \cite{ward1963hierarchical} for $t$-mixtures and swiss rolls, spectral clustering with self-tuning affinity \cite{zelnik2004self} for circles and moons. We benchmark across three SNR settings, which were adjusted such that the base estimator, informed with the correct number of clusters, achieves an ARI \cite{hubert1985comparing}, which we denote $\mathrm{ARI}(k^\star=5)$, of approximately $0.9$--$1.0$ (easy), $0.8$--$0.9$ (medium), and $0.7$--$0.8$ (hard). We generate datasets with $n=1500$ samples and $p=50$ features and $D=64$ embedded features for the RFF-embedded nonlinear shapes; we considered $B=20$ independent replicates per SNR level. We evaluate the benchmarked metrics across clustering solutions for $k \in \{3,4,5,6,7\}$. For each dataset we compared CARVE, under CARVE's \emph{max}, \emph{1-SE}, and \emph{quant} selection rules, against four CVIs --- Silhouette \cite{rousseeuw1987silhouettes}, Gap statistic \cite{tibshirani2001estimating}, Calinski--Harabasz \cite{calinski1974dendrite}, and Davies--Bouldin \cite{davies2009cluster} --- reporting mean $\mathrm{ARI}(\hat{k})$ against the true labels. Full simulation specifications (parameter settings, RFF construction, nuisance-dimension augmentation) are given in~\nameref{S3_Text}, and the complete overview of CARVE selectors, selection rules, and CVI adaptations is in~\nameref{S4_Text}.

\begin{figure}[!h]
    \centering
    \includegraphics[width=1\linewidth]{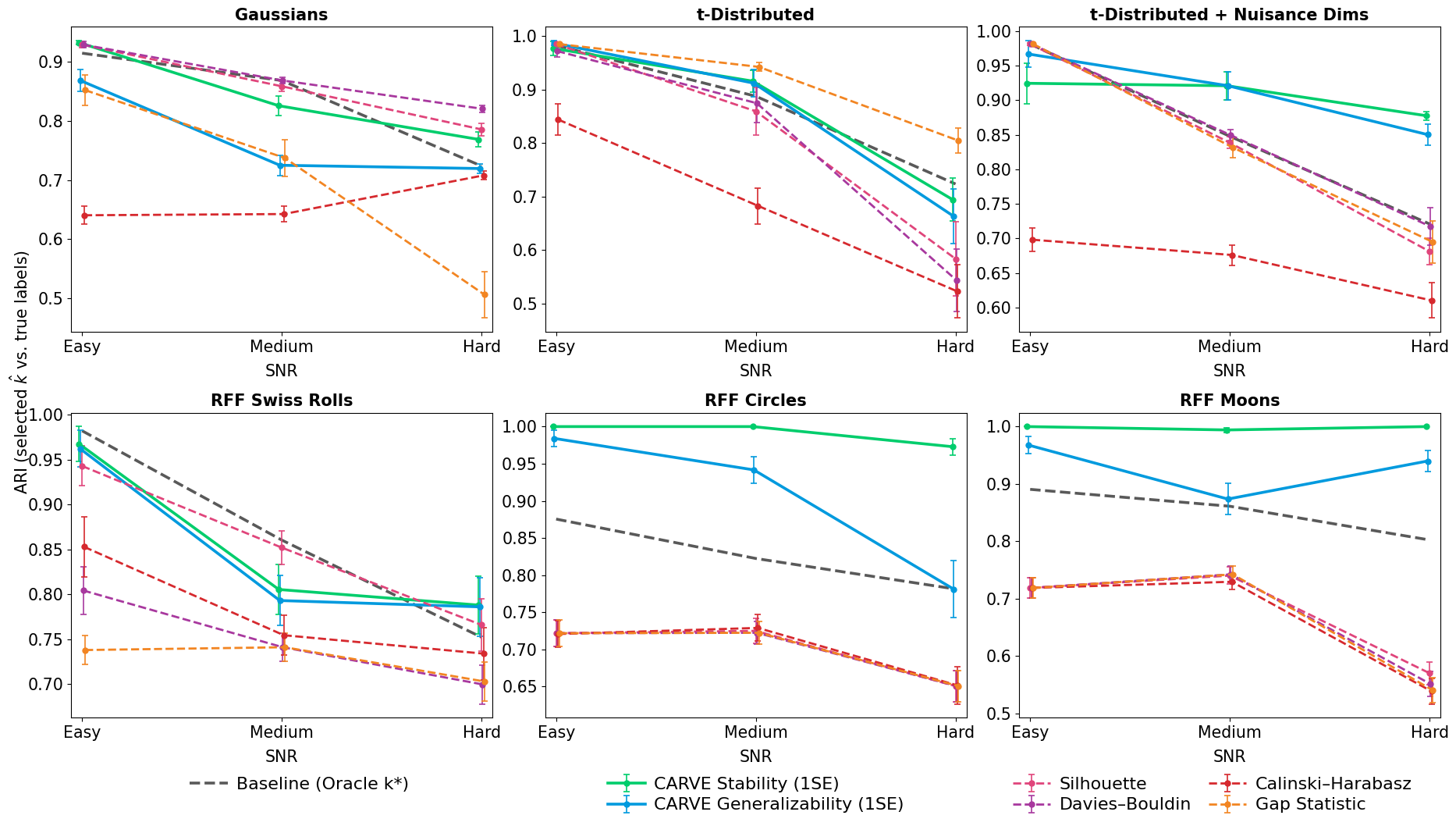}
    \caption{\textbf{Benchmarking performance across SNR levels} ($k^\star=5$).
    Mean $\mathrm{ARI}(\hat{k})$ at each of the three SNR settings -- easy (left), medium (middle), hard (right) -- for each clustering task. The grey line shows the oracle $\mathrm{ARI}(k^\star = 5)$; colored curves show CARVE (stability 1-SE rule and generalizability 1-SE rule), Silhouette, Gap statistic, Davies--Bouldin, and Calinski--Harabasz. Results are shown for $k^\star = 5$ with $B=20$ datasets per setting. Representative 2D PCA projections of each cluster shape at the hard setting are shown in \nameref{S1_Fig}.}
    \label{fig:benchmarking_results}
\end{figure}

\paragraph{Main findings.}
Across all six clustering tasks, performance varied with respect to the geometry of each setting (\textbf{Fig~\ref{fig:benchmarking_results}}). On Gaussian mixtures, most criteria were close to the oracle, and differences were modest. In heavy-tailed $t$-mixtures, CVIs such as Silhouette and Davies--Bouldin deteriorated at medium and hard SNR levels, while CARVE’s metrics maintained higher $\mathrm{ARI}(\hat{k})$. For $t$-mixtures with additional nuisance dimensions CARVE's metrics exceeded the oracle by selecting a more coherent partition at a nearby $k$ while CVIs performed poorly at medium and hard SNR levels. For swiss rolls, which are roughly spherical, CARVE's metrics provided advantages at easy and hard SNR settings while remaining competitive at medium settings. On circles and moons, CARVE’s metrics outperformed the CVIs consistently, while most CVIs struggled across all SNR levels. Calinski--Harabasz and the Gap statistic were the most consistently poor performers across clustering tasks. Representative 2D PCA projections of each cluster shape at the hard setting are shown in \nameref{S1_Fig}, and per-task tables in \nameref{S2_Table}--\nameref{S7_Table}.

Taken together, these experiments confirm that CARVE remained competitive or best in simple clustering tasks while offering advantages where geometry is nonlinear, distributions are heavy-tailed, or where noisy dimensions obscure the signal. CARVE’s stability metric with the 1-SE rule performed the most consistently strong across clustering tasks, while CARVE's generalizability metric with the 1-SE rule offered a supplementary perspective. Stability with the 1-SE rule is recommended and also implemented in our software as default. CVIs can be competitive in settings where geometry remains simple and spherical but should be interpreted with caution in more complex clustering tasks common in practice.

\paragraph{Scaling: accuracy and runtime.}
We additionally evaluated how $\mathrm{ARI}(\widehat{k})$ and runtime scale with sample size $n$ and feature dimension $p$ on Gaussian mixtures (clustering with KMeans). Stability remained stable across both axes. Generalizability degraded at medium-to-large $p$ because the default 100-tree random forest is unable to capture sufficient structure in higher dimensions; we recommend increasing the number of trees, or supplying a different classifier, as $p$ grows. Runtime is dominated by the cost of the clustering algorithm and the classifier, with generalizability being slower than stability on average. Full experimental setup, accuracy and runtime curves, and per-axis tables are provided in \nameref{S5_Text} (\nameref{S2_Fig}, \nameref{S3_Fig}, \nameref{S8_Table}--\nameref{S10_Table}).

\subsection*{Case studies}
We present case studies that illustrate how clustering validation directly affects biological interpretation and how CARVE can improve reproducibility. Both examples represent situations in which clustering may be used as foundation for downstream scientific claims---identifying cell states during stem-cell differentiation \cite{klein2015droplet} or discovering prognostically relevant cell populations in leukemia \cite{levine2015data}. Selecting a suboptimal clustering solution may risk missing biologically meaningful structure or reporting dubious subpopulations. The two datasets also span different technical challenges: a droplet-based scRNA-seq experiment with transcriptionally continuous populations, and a mass cytometry benchmark whose heavy-tailed protein expression distributions and overlapping cell populations challenge geometry-based CVIs.

\paragraph{Case study 1: Droplet-based scRNA-seq.}
Klein et al.\ \cite{klein2015droplet} used droplet-based scRNA-seq to dissect the heterogeneous differentiation of mouse embryonic stem cells (mESCs) after withdrawal of leukemia inhibitory factor (LIF), with transcriptional states at successive time points (d0, d2, d4, d7). We analyzed a preprocessed subsample of 1{,}358 cells across 2{,}000 highly variable genes, labeled by time points; full preprocessing is detailed in \nameref{S6_Text}.

\begin{figure}[!h]
    \centering
    \includegraphics[width=1\linewidth]{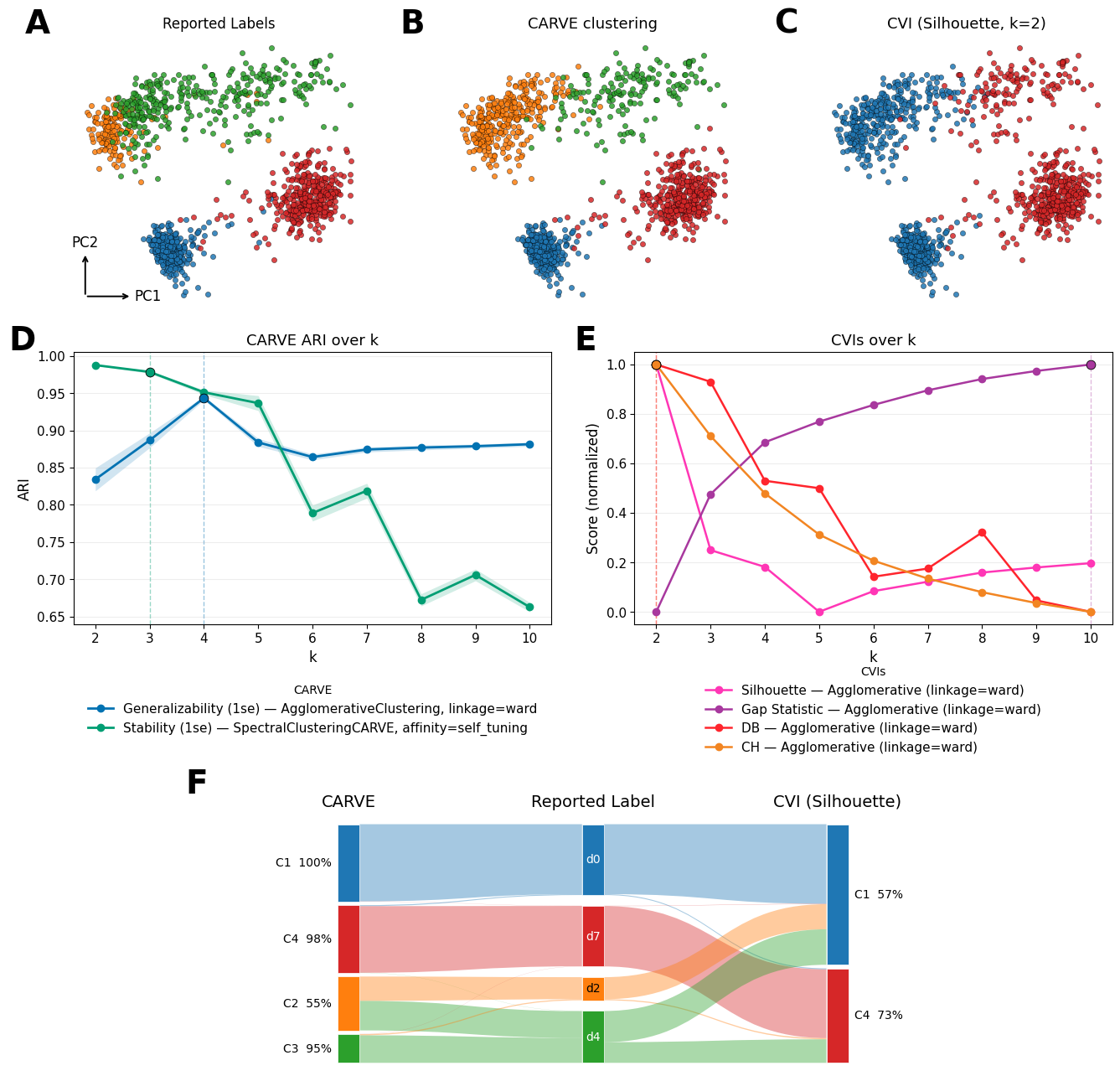}
    \caption{\textbf{Klein droplet-based scRNA-seq: overview and CARVE vs.\ Silhouette comparison.}
    \textbf{(A)} PCA of the preprocessed gene expression matrix, with cells colored by reported stages.
    \textbf{(B)} PCA embedding colored by
    consensus clustering labels derived from the Ward agglomerative consensus matrix at $k=4$.
    \textbf{(C)} Clustering selected by maximizing the Silhouette statistic, visualized on the PCA embedding.
    \textbf{(D)} CARVE's generalizability and stability ARIs as functions of $k$, with the selected base model highlighted using the 1-SE rule.
    \textbf{(E)} Silhouette, Gap statistic, Davies--Bouldin (DB), and Calinski--Harabasz (CH) evaluated across $k$ for multiple clustering models and hyperparameters.
    \textbf{(F)} Alluvial plot comparing CARVE consensus clusters (left) and Silhouette-selected clusters (right) through reference cell-type labels (center). 
    }
    \label{fig:klein_results}
\end{figure}

\paragraph{CARVE results.}
We ran CARVE across Ward agglomerative and spectral clustering with self-tuning affinity for $k \in \{2,\dots,10\}$. The generalizability ARI with the 1-SE rule selected Ward agglomerative clustering at $k=4$ (\textbf{Fig~\ref{fig:klein_results}(C)}). While the stability ARI did not indicate a clear preference, CARVE's exploration and visualization tools (see \textbf{Fig~\ref{fig:carve_output_klein}}) supported Ward agglomerative clustering at $k=4$ as the preferred solution (\textbf{Fig~\ref{fig:klein_results}(B, D)}). The $k=4$ solution coincides with the four sampled differentiation stages after LIF withdrawal, recovering the transcriptional progression described by Klein et al. In contrast, three CVIs (Silhouette, CH, DB) maximized at $k=2$ under Ward agglomerative clustering, yielding a coarse partition that merges reported stages (\textbf{Fig~\ref{fig:klein_results}(C, E, F)}); an analysis only relying on CVIs (as is common in practice) would have collapsed the d0, d2 and d4, d7 transcriptional states and missed the continuity between stages d2 and d4 entirely. The full per-criterion breakdown --- CARVE and the four CVIs evaluated across Ward agglomerative and spectral clustering --- is in \nameref{S6_Text}.

\paragraph{Case study 2: Mass Cytometry.}
Levine et al.\ \cite{levine2015data} performed a data-driven phenotypic dissection of acute myeloid leukemia using mass cytometry. We analyzed the Levine 32-dimensional CyTOF benchmark, consisting of a stratified subsample of 5{,}000 cells across 32 ``type'' markers drawn from 104{,}184 labeled bone-marrow cells spanning 14 manually gated populations; full preprocessing is detailed in \nameref{S7_Text}.

\begin{figure}[!h]
    \centering
    \includegraphics[width=1\linewidth]{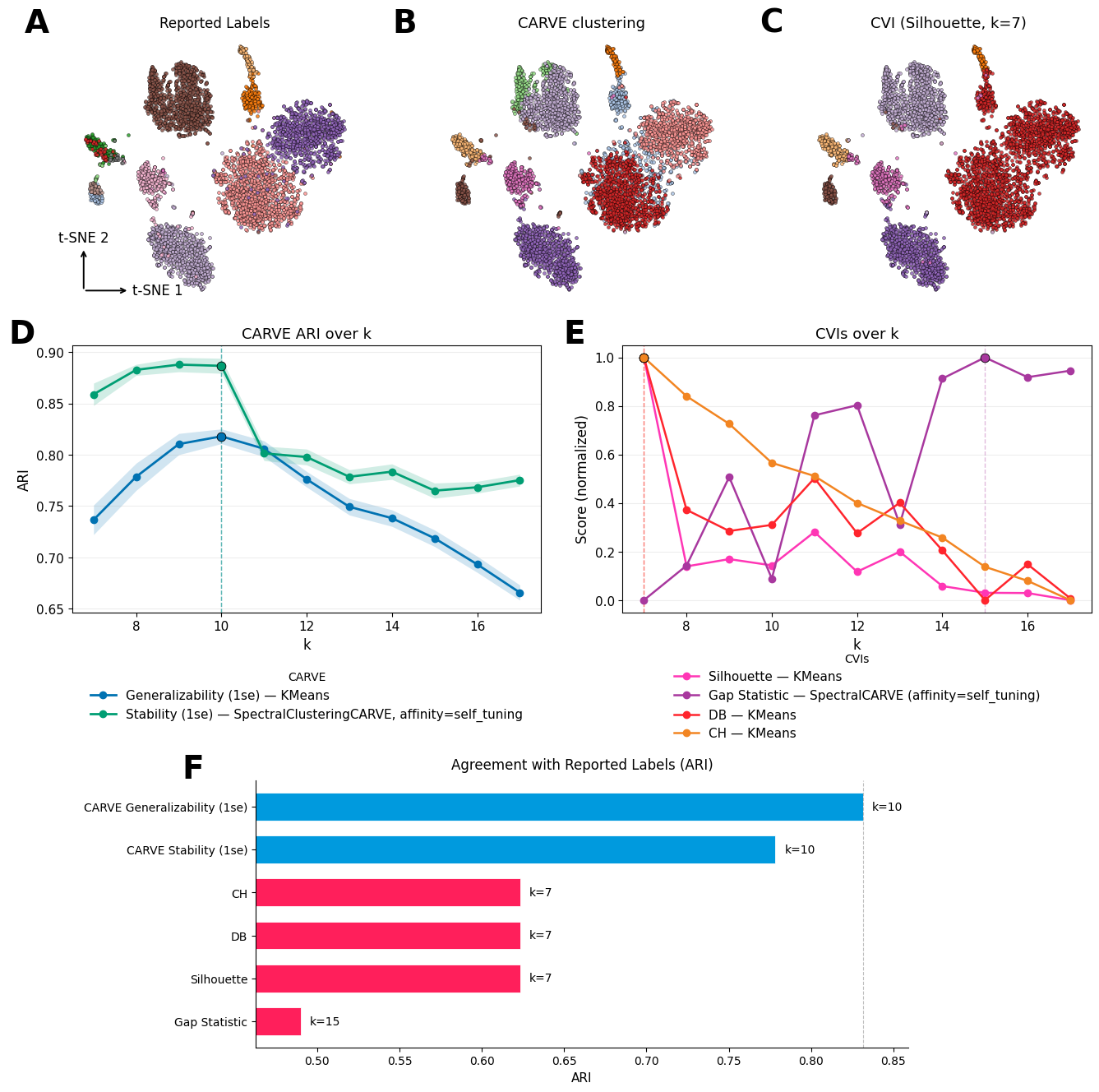}
    \caption{\textbf{Levine 32dim mass cytometry: overview and CARVE vs.\ CVI comparison.}
    \textbf{(A)} t-SNE \cite{vandermaaten2008visualizing} embedding of the preprocessed marker matrix, with cells colored by the 14 manually gated population labels.
    \textbf{(B)} t-SNE embedding colored by the CARVE-selected clustering at $k=10$.
    \textbf{(C)} t-SNE embedding colored by the Silhouette-selected clustering (KMeans, $k=7$).
    \textbf{(D)} CARVE internal ARI as a function of $k$: generalizability (KMeans, 1-SE) and stability (spectral clustering with self-tuning affinity, 1-SE), with standard-error bands.
    \textbf{(E)} CVIs (Silhouette, Gap statistic, Davies--Bouldin, Calinski--Harabasz) as a function of $k$.
    \textbf{(F)} Agreement with reported labels (ARI) for each selection method; CARVE generalizability and stability at $k=10$ substantially outperform all CVIs.}
    \label{fig:levine_results}
\end{figure}

\paragraph{CARVE results.}
We ran CARVE with KMeans and spectral clustering with self-tuning affinity for $k \in \{7,\dots,17\}$. Both the stability and generalizability 1-SE rules selected $k=10$: CARVE's stability formed a broad plateau over $k=8$--$10$ and preferred spectral clustering, while CARVE's generalizability peaked for KMeans at $k=10$ (\textbf{Fig~\ref{fig:levine_results}(D)}). Against the 14 reported labels the CARVE-selected partitions achieve $\mathrm{ARI} \approx 0.84$ (generalizability) and $\mathrm{ARI} \approx 0.78$ (stability), substantially exceeding all CVI selected clustering solutions ($\mathrm{ARI} \le 0.63$; Silhouette, CH, and DB suggest KMeans at $k=7$, while the gap statistic selects spectral clustering at $k=15$; \textbf{Fig~\ref{fig:levine_results}(E, F)}). The $k=10$ solution recovers more of the population structure visible in the reported labels while also being more stable to perturbations and generalizable on held-out data than the coarse $k=7$ clustering indicated by the CVIs (\textbf{Fig~\ref{fig:levine_results}(D)}). Upon further examination (see \nameref{S7_Text}), the coarser CVI-selected $k=7$ partition collapses several T- and NK-cells into a single cluster -- grouping CD4$^+$ T, CD8$^+$ T, and CD16$^-$ NK cells together -- whereas CARVE's $k=10$ solution separates these as distinct groups, more faithfully recovering the reported populations \cite{levine2015data}. In-detail overviews are located in \nameref{S7_Text} and CARVE's output can be seen in \nameref{S4_Fig}.

\paragraph{Case study summary.}
CARVE's multi-level analysis delivers stronger analytical capabilities than existing CVI-based methods. On the Klein data, CARVE identified continuous transitional cell states at the cluster level (\textbf{Fig~\ref{fig:carve_output_klein}(D, F)}), whereas most CVIs collapsed many stages into a single coarse cluster (\textbf{Fig~\ref{fig:klein_results}(C)}). On the Levine data, CARVE recovered clustering solutions whose ARI against the reported labels substantially exceeds every compared method. Across both case studies, CARVE outperforms existing validation methods not only by agreement with reported labels but, more importantly, by enabling flexible, multi-layered comparison of multiple clustering solutions.

\section*{Discussion}
Clustering validation is, in current practice, severely underserved: most clustering analyses report a single solution without any form of validation, and the analyses that do attempt validation overwhelmingly rely on classical CVIs, whose geometric and distributional assumptions are rarely met by the data sets encountered in genomics and other biomedical fields. We have presented CARVE, a resampling-based framework that unifies stability and generalizability-based clustering validation into a single, reproducible workflow. Our results show that CARVE recovers near optimal clusterings across a broad range of simulation environments --- Gaussian and heavy-tailed mixtures, settings with nuisance dimensions, and nonlinear manifolds. On real biomedical data, CARVE recovered a four-stage mESC differentiation trajectory \cite{klein2015droplet} and a fine-grained bone-marrow population structure \cite{levine2015data} --- biological findings that the CVI-preferred partitions collapsed and missed entirely.

Several pathways for future work remain for the broader clustering-validation community. The most pressing direction is the validation of clustering on multimodal data. As single-cell experiments increasingly couple transcriptomic, epigenetic, and proteomic modalities \cite{baysoy2023technological}, open questions remain on how to construct stability and generalizability measures across modalities and across data integration pipelines.

Existing tools only cover individual pieces of clustering validation workflows --- stability scores, consensus matrices, individual CVIs --- leaving practitioners to assemble ad hoc pipelines from independently developed components. CARVE brings resampling stability, held-out generalizability, multi-level diagnostics at the global, cluster, and sample level, principled selection rules, and consensus-based cluster labels together within a single scikit-learn-compatible Python API, with an analogous R interface that supports integration into existing workflows. By making this collection of validation capabilities accessible to data scientists and practitioners alike, CARVE offers a practical path toward improving the trust, replicability, and reliability of clustering results in computational biology and beyond.

\section*{Acknowledgments}
The authors acknowledge funding from NSF DMS-2516872.

\section*{Software Availability}

CARVE is released as an open-source Python package, \texttt{carve}, distributed under the MIT license. It exposes a scikit-learn-compatible API and is installable from the Python Package Index with \texttt{pip install carve-validate} (Python 3.12). The source code, usage documentation, and tutorial notebooks are hosted on GitHub at \url{https://github.com/DataSlingers/CARVE}, along with all benchmarking code and case-study notebooks used in this manuscript to support reproducibility \cite{wycik_2026_20448965}. A companion R package provides an analogous interface with native support for \texttt{Seurat} and \texttt{SingleCellExperiment} workflows; it is available from the same repository and can be installed with \texttt{remotes::install\_github("DataSlingers/CARVE", subdir = "code/carve-r")}.

\newpage

\section*{Supporting information}


\paragraph*{S1 Text.}
\label{S1_Text}
\textbf{Detailed methodology, pseudocode, and internal design of CARVE.}

\textbf{Detailed methodology.}
For each candidate clustering configuration (a clustering estimator \(f\) with hyperparameters \(\theta\), including the cluster number \(k\)), CARVE quantifies two properties across \(B\) resampling iterations: \emph{stability} (agreement of cluster assignments under perturbations of the data) and \emph{generalizability} (agreement between a clustering on held-out samples and labels predicted from an in-sample clustering model).

For a fixed configuration \((f,\theta)\), each iteration \(b \in \{1,\dots,B\}\) proceeds as follows. If randomized preprocessing is enabled, CARVE samples a preprocessing pipeline \(T^{(b)}\) by drawing one normalization transform and one dimensionality-reduction transform from user-provided option sets; otherwise \(T^{(b)}\) is the identity. The pipeline \(T^{(b)}\) is fit on \(X\) yielding a preprocessed representation \(\widetilde{X}^{(b)} = T^{(b)}(X)\).
Then, CARVE draws two independent subsamples \(P_1^{(b)}\) and \(P_2^{(b)}\) of size approximately \(\lfloor \rho n \rfloor\) from \(\widetilde{X}^{(b)}\), and defines a holdout set \(P_{\mathrm{test}}^{(b)}\) as the complement of \(P_1^{(b)}\).

Next, CARVE applies the clustering method \(f(\cdot;\theta)\) to \(\widetilde{X}^{(b)}\) restricted to each index set, producing cluster labelings
\(C_1^{(b)} = f(\widetilde{X}^{(b)}[P_1^{(b)}];\theta)\),
\(C_2^{(b)} = f(\widetilde{X}^{(b)}[P_2^{(b)}];\theta)\), and
\(C_{\mathrm{test}}^{(b)} = f(\widetilde{X}^{(b)}[P_{\mathrm{test}}^{(b)}];\theta)\).
Stability for iteration \(b\) is defined as the adjusted Rand index (ARI) \cite{hubert1985comparing} between the two subsample clusterings evaluated on their intersection:
\[
S^{(b)} = \mathrm{ARI}\!\left(C_1^{(b)}\big|_{P_1^{(b)}\cap P_2^{(b)}},\; C_2^{(b)}\big|_{P_1^{(b)}\cap P_2^{(b)}}\right).
\]
Generalizability for iteration \(b\) is computed by training a classifier \(h^{(b)}\) on \((\widetilde{X}^{(b)}[P_1^{(b)}], C_1^{(b)})\), predicting labels on the holdout set \(\widehat{C}_{\mathrm{test}}^{(b)} = h^{(b)}(\widetilde{X}^{(b)}[P_{\mathrm{test}}^{(b)}])\), and evaluating
\[
G^{(b)} = \mathrm{ARI}\!\left(C_{\mathrm{test}}^{(b)},\; \widehat{C}_{\mathrm{test}}^{(b)}\right).
\]
In the current implementation, \(h^{(b)}\) is a random forest classifier \cite{breiman2001random} with 100 trees by default; however, users may easily provide their own classifiers. Note that the ARI is permutation-invariant, so label identities need not be aligned for the ARI-based scores.

Across iterations, CARVE summarizes \(\{S^{(b)}\}_{b=1}^B\) and \(\{G^{(b)}\}_{b=1}^B\) into means, standard errors, and 95\% confidence intervals. In addition, CARVE constructs a consensus matrix \(M\) from the subsample-label-pairs \(\{(P_1^{(b)}, C_1^{(b)})\}_{b=1}^B\), where each entry \(M_{ij}\) is the fraction of times samples \(i\) and \(j\) were assigned to the same cluster among the iterations in which they were co-sampled; per-sample stability scores are derived from the consensus matrix as Gini and cross-entropy of the respective rows which may then also be aggregated cluster-wise. We also derive from the consensus matrices the proportion of ambiguous clusters (PAC) \cite{senbabaouglu2014critical}. Finally, CARVE computes a sample-level generalizability array \(E \in [0,1]^n\) from the holdout predictions by recording, for each sample, the fraction of iterations in which its predicted label matches the holdout clustering label after aligning predicted labels to the holdout labels; the mean of \(E\) yields a global out-of-sample prediction accuracy (equivalently \(1\) minus an average misclassification rate) for the configuration; aggregating \(\{E_i\}_{i=1}^n\) cluster-wise yields cluster-level accuracy scores.

\textbf{Pseudocode.}
We write ARI for the adjusted Rand index \cite{hubert1985comparing} and we use \(\rho\in(0,1)\) for the subsampling proportion. Each configuration \(c\) corresponds to a clustering estimator \(f\) with hyperparameters \(\theta\), including the cluster number \(k\). 

\begin{algorithm}[h!]
\caption{CARVE validation for a fixed configuration $(f,k)$}
\label{alg:carve_validation}
\DontPrintSemicolon

\begingroup
\setstretch{1}
\footnotesize
\SetAlgoSkip{0.5em}
\SetAlgoNlRelativeSize{-1}

\KwIn{
Data matrix $X \in \mathbb{R}^{n\times p}$;
clustering estimator $f$ with $k$ clusters;
resampling iterations $B$;
subsampling proportion $\rho \in (0,1)$;
label-prediction model $\hat{h}$
}
\KwOut{
$\{S_b\}_{b=1}^B$ (ARI stability), $\{G_b\}_{b=1}^B$ (ARI generalizability),
consensus matrix $\widehat{M}_{f,k}$,
per-sample misclassification rates $\widehat{E}_{f,k}(i)$
}

Initialize co-clustering counts $M_{f,k}(i,j)\gets 0$ and co-sampling counts $W_{f,k}(i,j)\gets 0$ for all $1\le i<j\le n$\;
Initialize misclassification counts $E_{f,k}(i)\gets 0$ and test-set counts $V_{f,k}(i)\gets 0$ for all $i\in\{1,\dots,n\}$\;

\For{$b \gets 1$ \KwTo $B$}{
    Sample $P_1^{(b)}, P_2^{(b)} \subseteq \{1,\dots,n\}$ independently with $|P_1^{(b)}|=|P_2^{(b)}|=\lfloor \rho n \rfloor$\;
    Set $P_{\mathrm{test}}^{(b)} \gets \{1,\dots,n\}\setminus P_1^{(b)}$\;

    $C_1^{(b)} \gets f(X[P_1^{(b)}])$, \;
    $C_2^{(b)} \gets f(X[P_2^{(b)}])$, \;
    $C_{\mathrm{test}}^{(b)} \gets f(X[P_{\mathrm{test}}^{(b)}])$\;

    $S_b \gets \mathrm{ARI}\!\left(C_1^{(b)}|_{P_1^{(b)} \cap P_2^{(b)}},\, C_2^{(b)}|_{P_1^{(b)} \cap P_2^{(b)}}\right)$\;

    Update $(M_{f,k},W_{f,k})$ using $P_1^{(b)}$ and labels $C_1^{(b)}$;

    Train $\hat{h}^{(b)}$ on $\left(X[P_1^{(b)}],\, C_1^{(b)}\right)$\;
    $\widehat{C}_{\mathrm{test}}^{(b)} \gets \hat{h}^{(b)}(X[P_{\mathrm{test}}^{(b)}])$\;
    $G_b \gets \mathrm{ARI}\!\left(C_{\mathrm{test}}^{(b)},\, \widehat{C}_{\mathrm{test}}^{(b)}\right)$\;

    Compute a label matching $\sigma^{(b)}$ between $\widehat{C}_{\mathrm{test}}^{(b)}$ and $C_{\mathrm{test}}^{(b)}$ (Hungarian assignment \cite{kuhn1955hungarian})\;
    \ForEach{$i \in P_{\mathrm{test}}^{(b)}$}{
        $V_{f,k}(i) \gets V_{f,k}(i) + 1$\;
        \If{$C_{\mathrm{test}}^{(b)}(i) \neq \sigma^{(b)}(\widehat{C}_{\mathrm{test}}^{(b)}(i))$}{
            $E_{f,k}(i) \gets E_{f,k}(i) + 1$\;
        }
    }
}

Define $\widehat{M}_{f,k}(i,j) \gets M_{f,k}(i,j)/W_{f,k}(i,j)$ when $W_{f,k}(i,j)>0$ (else $\mathrm{NA}$)\;
Define $\widehat{E}_{f,k}(i) \gets E_{f,k}(i)/V_{f,k}(i)$ when $V_{f,k}(i)>0$ (else $\mathrm{NA}$)\;

\endgroup
\end{algorithm}

\textbf{Implementation and default configuration.}
CARVE is implemented as a Python library whose interface follows the scikit-learn estimator API and as a R package which integrates into Seurat workflows. The user first instantiates a \texttt{CARVE()} estimator and then runs the analysis via \texttt{fit()}. If no parameters are supplied at instantiation, CARVE runs with a default configuration (\nameref{S1_Table}). CARVE includes a custom spectral clustering implementation (\texttt{SpectralClusteringCARVE}) that supports self-tuning local scaling \cite{zelnik2004self}, RBF, and $k$-nearest-neighbor affinity construction; the default configuration uses the self-tuning affinity, which adapts the kernel scale to the local density around each point and requires no global bandwidth parameter (see \nameref{S1_Table}). Users may also override the default options by supplying custom grids for clustering estimators and preprocessing steps (normalization and dimensionality reduction). Clustering estimator grids are specified as a list of \texttt{(EstimatorClass, param\_grid)} tuples, where \texttt{param\_grid} maps parameter names to explicit candidate lists; for each estimator, CARVE evaluates the Cartesian product of candidates within that estimator's grid. Preprocessing options are provided as collections of candidate transforms from which CARVE samples during resampling. 

\paragraph*{S1 Table.}
\label{S1_Table}
\textbf{Default configuration for the \texttt{CARVE} estimator.}
Some defaults are defined algorithmically and depend on the input data matrix $X \in \mathbb{R}^{n \times p}$ and the subsampling proportion $\rho$.

\begin{table}[!ht]
\centering
\caption{\textbf{Default parameters and data-dependent default option sets.}}
\resizebox{\textwidth}{!}{%
\begin{tabular}{|l|l|p{3.2in}|}
\hline
\textbf{Parameter} & \textbf{Default} & \textbf{Definition / notes} \\
\hline
\texttt{n\_clusters} & $\{2,\dots,10\}$ & Cluster counts to evaluate. Accepts an integer \texttt{K} (expanded to $\{2,\dots,K\}$) or an explicit array. \\
\hline
\texttt{n\_resamples} & $100$ & Number of resampling iterations $B$. \\
\hline
\texttt{subsample\_ratio} & $0.618$ & Subsampling proportion $\rho$. Each training subsample has size approximately $\lfloor \rho n \rfloor$. \\
\hline
\texttt{estimator\_param\_grids} & \texttt{"light"} &
Either a preset string (\texttt{"light"} or \texttt{"full"}) or a user-supplied list of \texttt{(EstimatorClass, param\_grid)} tuples. The \texttt{light} preset (default) includes:
(i) \texttt{KMeans} with \texttt{n\_clusters}$\in\{2,\dots,K\}$;
(ii) \texttt{AgglomerativeClustering} with \texttt{n\_clusters}$\in\{2,\dots,K\}$ and \texttt{linkage=ward};
(iii) \texttt{SpectralClusteringCARVE} with \texttt{n\_clusters}$\in\{2,\dots,K\}$ and \texttt{affinity=self\_tuning} \cite{zelnik2004self}.
The \texttt{full} preset additionally includes agglomerative clustering with average, single, and complete linkage, and \texttt{SpectralClusteringCARVE} with \texttt{affinity=rbf} and data-driven gamma values estimated via a $k$-NN median heuristic ($k=7$) with multipliers $(0.5, 1.0, 2.0)$. \\
\hline
\texttt{normalization\_options} & \texttt{None} &
If \texttt{None}, CARVE uses \texttt{default\_normalization\_options()}:
(i) identity transform;
(ii) \texttt{StandardScaler} (zero mean, unit variance);
(iii) \texttt{log1p} transform. \\
\hline
\texttt{dim\_reduction\_options} & \texttt{None} &
If \texttt{None}, CARVE uses \texttt{default\_dim\_reduction\_options(X, subsample\_ratio)}.
Let $\texttt{min\_n}=\operatorname{round}\!\bigl((1-\rho)\,n\bigr) - 1$.
Options include:
(i) identity transform;
(ii) \texttt{PCA} with \texttt{n\_components}$\in\{2,\dots,\min(\texttt{min\_n},p)-1\}$;
(iii) \texttt{tSNE} with \texttt{n\_components}$=2$ and \texttt{perplexity}$\in\{5,\dots,\min(\texttt{min\_n},51)-1\}$;
(iv) \texttt{UMAP} with \texttt{n\_neighbors}$\in\{5,\dots,50\}$, \texttt{min\_dist}$=0.1$, and \texttt{n\_components}$\in\{2,\dots,\min(\texttt{min\_n},p)-1\}$. \\
\hline
\texttt{classifier} & \texttt{None} & Classifier used to score generalizability via held-out prediction. If \texttt{None}, CARVE uses \texttt{RandomForestClassifier} from scikit-learn. \\
\hline
\texttt{n\_trees} & $100$ & Number of trees passed to the default \texttt{RandomForestClassifier} when \texttt{classifier} is \texttt{None}. \\
\hline
\texttt{reference\_labels} & \texttt{None} & Optional reference labels used for consistent plotting and label alignment when supplied. \\
\hline
\texttt{n\_jobs} & $1$ & Parallelism (number of workers). \\
\hline
\texttt{random\_state} & \texttt{None} & RNG seed; if set, results are reproducible conditional on the computational environment. \\
\hline
\texttt{verbose} & $0$ & Logging verbosity. \\
\hline
\end{tabular}%
}
\label{tab:S1_defaults}
\end{table}

\paragraph*{S2 Text.}
\label{S2_Text}
\textbf{User input and output}
\paragraph{Input.} Users provide the \texttt{fit()} method with a data matrix \(X \in \mathbb{R}^{n \times p}\), where \(n\) is the number of samples and \(p\) is the number of features. Optionally, users may provide reference labels \texttt{reference\_labels}. These labels are used as a reference to keep cluster label identities consistent across plots, by aligning new labelings to the stored reference. 

While running, CARVE prints progress output: a header that summarizes the run configuration (including \(K\), \(B\), \(\rho\), the evaluated estimator grid, and parallelization settings), and a footer after completion. 

\paragraph{Output.} CARVE exposes results as tables and consensus objects, most importantly \texttt{model\_df\_} (per-configuration metrics across estimators and \(k\)) and, when randomized preprocessing is enabled, \texttt{pipeline\_df\_} (metrics stratified by preprocessing pipeline). In addition to ARI-based stability/generalizability, CARVE also computes consensus-based stability scores (PAC, Gini, CE) and a global accuracy-based generalizability score. 

\paragraph{Example of user-specified option sets.}
Clustering estimators are provided as a list of \texttt{(EstimatorClass, param\_grid)} tuples; CARVE enumerates the Cartesian product of candidates within each estimator's \texttt{param\_grid}.
Preprocessing options (normalization and dimensionality reduction) are provided as collections of candidate transforms from which CARVE samples stochastically during resampling.

\begin{verbatim}
# Example: user-defined estimator grids (exhaustive within each estimator grid)
estimator_param_grids = [
    (KMeans, {"n_clusters": [2, 3, 4, 5, 6]}),
    (AgglomerativeClustering, {
        "n_clusters": [2, 3, 4, 5, 6],
        "linkage": ["ward", "complete", "single", "average"]
    }),
    (SpectralClusteringCARVE, {
        "n_clusters": [2, 3, 4, 5, 6],
        "affinity": ["self_tuning"]
    })
]
\end{verbatim}

\paragraph*{S3 Text.}
\label{S3_Text}
\textbf{Simulation environment specifications.}
We describe the simulation procedure and parameter settings for each of the six benchmarking clustering tasks.
All clustering tasks share: $n = 1500$ total samples, $p = 50$ features for Gaussians and t-distributed data, $D = 64$ embedding dimensions for RFF embedded nonlinear shapes, $k^\star = 5$ true clusters, and $B = 20$ independent datasets per SNR setting.
Candidate cluster counts are $k \in \{3,4,5,6,7\}$.

\textbf{SNR settings.}
Each clustering task is parameterized by three SNR settings --- easy, medium, and hard --- whose parameter settings were calibrated so that a base estimator (informed with the true $k^\star = 5$) achieves a target mean $\mathrm{ARI}(k^\star = 5)$ against the true labels in approximately $[0.9, 1.0]$ (easy), $[0.8, 0.9]$ (medium), and $[0.7, 0.8]$ (hard). Calibration was performed on $B_{\text{datasets}} = 20$ seeds drawn with the same random state used during the actual benchmarking, so that the calibration and benchmarking data sets coincide.

The mechanisms below are shared across multiple clustering tasks and are described once here.

\textbf{Centroid placement.}
Cluster centroids are placed using a best-candidate algorithm. The first centroid is drawn uniformly from the hypercube $[-c,\,c]^{p}$, where $c = \texttt{center\_box} = 3.0$. For each subsequent centroid, $n_{\mathrm{cand}} = \texttt{n\_candidates} = 64$ random candidate points are drawn uniformly from the same hypercube; the candidate whose minimum squared-Euclidean distance to all already-placed centroids is largest is retained. Formally, if $\mathcal{C}_{i-1} = \{c_1,\dots,c_{i-1}\}$ are the centroids already placed, the $i$-th centroid is

\[
  c_i = \arg\max_{q \,\in\, \mathcal{Q}} \;\min_{c \,\in\, \mathcal{C}_{i-1}} \|q - c\|^2,
  \qquad \mathcal{Q} \subset [-c,c]^p,\; |\mathcal{Q}| = 64.
\]

This heuristic avoids degenerate configurations in which two true cluster centroids are nearly coincident, and produces well-spread centroid configurations without requiring a minimum-distance hard constraint that could fail for large $k^\star$ or small $p$.

\textbf{Cluster scale and correlation.} The parameter \texttt{cluster\_scale} $= s_c \ge 0$ specifies the scale of cluster $c$. Internally, each cluster's covariance matrix is $\Sigma_c = s_c^2\, R$, where $R$ is a $p\times p$ AR(1) correlation matrix. For Gaussian clusters, samples are drawn from $\mathcal{N}(\mu_c, \Sigma_c)$; for $t$-distributed clusters the same $\Sigma_c$ enters the Cholesky parameterisation of the multivariate $t$. Larger $s_c$ produces a more diffuse (harder to separate) cluster; unequal $s_c$ across clusters produces heteroscedastic geometries. The correlation matrix $R \in \mathbb{R}^{p\times p}$ has entries
  
\[
    R_{ij} = \rho^{|i-j|}, \qquad |\rho| < 1,
\]

where $\rho = \texttt{corr\_strength}$, so adjacent features have correlation $\rho$ and correlation decays geometrically with lag. The resulting cluster covariance is $\Sigma_c = s_c^2 R$. When \texttt{corr\_type='none'}, $R = I_p$ (independent features).

\textbf{Cluster-size sampling.} Given $k^\star$ clusters and $n = 1500$ total samples, a per-cluster floor $f = \max(5,\,\lceil 0.1\,n/k^\star \rceil)$ is first guaranteed to every cluster (ensuring no cluster is too small or empty). The remaining $n - f\,k^\star$ samples are then distributed across clusters via a Dirichlet--Multinomial draw:
  
\[
    \pi \sim \mathrm{Dirichlet}(\alpha,\dots,\alpha),
    \qquad
    \Delta n_c \sim \mathrm{Multinomial}(n - fk^\star,\; \pi),
    \qquad
    n_c = f + \Delta n_c.
\]

The concentration $\alpha = \texttt{cluster\_size\_dirichlet\_alpha}$ controls imbalance: small $\alpha$ (e.g.\ $0.1$) yields highly skewed sizes; large $\alpha$ (e.g.\ $0.9$) yields near-balanced sizes.

\textbf{Cluster shape 1: Isotropic Gaussian mixtures.}
\texttt{distribution='gaussian'}, \texttt{corr\_type='none'}; base estimator: K-means.
SNR is governed by increasing cluster spread (\texttt{cluster\_scale}: $4.0 / 4.5 / 4.6$ from easy to hard) and growing cluster-size imbalance (Dirichlet concentration $\alpha$: $0.9 / 0.5 / 0.1$).

\textbf{Cluster shape 2: Heavy-tailed $t$-distributed mixtures.}
\texttt{distribution='t'}, \texttt{corr\_type='none'}; base estimator: Ward agglomerative clustering.
SNR is governed jointly by heavier tails (degrees of freedom $\nu$: $5$ at easy, $3$ at medium and hard), larger cluster-scale heterogeneity (\texttt{cluster\_scale}: $[3,1,1,1,1]$ at easy, $[3,1.5,1,1,1]$ at medium, $[4,3,2,1,1]$ at hard), and increasing size imbalance ($\alpha$: $0.9 / 0.3 / 0.1$).

\textbf{Cluster shape 3: $t$-distributed mixtures with high-dimensional nuisance features.}
\texttt{distribution='t'}, \texttt{corr\_type='ar1'}; base estimator: Ward agglomerative clustering.
Signal features ($p=50$) remain fixed; additional noise dimensions are drawn from a $t$-distribution with scale matched to the signal features (\texttt{noise\_scale='match'}).
SNR is governed by increasing noise dimensionality (\texttt{noise\_dims}: $512 / 1280 / 1536$), growing feature correlation (\texttt{corr\_strength}: $0.1 / 0.3 / 0.5$), heavier tails ($\nu$: $5 / 4 / 3$), and increasing size imbalance ($\alpha$: $0.9 / 0.3 / 0.1$); cluster spread stays close to $1.0$ (with one cluster slightly enlarged at medium and hard).

\textbf{Cluster shapes 4--6: Nonlinearly structured shapes (circles, moons, Swiss rolls) via random Fourier features.}
\texttt{distribution} $\in$ \{\texttt{'circles'}, \texttt{'moons'}, \texttt{'swiss\_roll'}\}, \texttt{embed\_dim=64}, \texttt{corr\_type='ar1'}; base estimator: spectral clustering with self-tuning affinity \cite{zelnik2004self} for circles and moons, Ward agglomerative clustering for Swiss rolls.
Cluster manifolds are generated in $\mathbb{R}^2$, AR(1) correlation is applied, the result is mapped to $\mathbb{R}^{64}$ via random Fourier features (RFF).
The RFF map is
\[
\phi(x) = \sqrt{\tfrac{2}{D}}\cos(Wx+b),\quad
W_{ij}\overset{\mathrm{iid}}{\sim}\mathcal{N}(0,\sigma^{-2}),\quad
b_j\overset{\mathrm{iid}}{\sim}\mathrm{Unif}(0,2\pi),
\]
where $\sigma = \texttt{embed\_param}$ is the kernel lengthscale.
The SNR parameters are task-specific:
circles use \texttt{embed\_param}: $12.0 / 7.3 / 6.0$, $\alpha$: $0.9 / 0.61 / 0.35$, and \texttt{corr\_strength}: $0.1 / 0.23 / 0.20$;
moons use \texttt{embed\_param}: $10.7 / 5.7 / 14.5$ (lengthscale is non-monotone --- the hard setting combines a longer lengthscale with stronger overlap from cluster-scale and imbalance), $\alpha$: $0.67 / 0.57 / 0.10$, and \texttt{corr\_strength}: $0.39 / 0.30 / 0.16$;
Swiss rolls use \texttt{embed\_param}: $8.0 / 5.0 / 5.0$, $\alpha$: $0.9 / 0.7 / 0.5$, and \texttt{corr\_strength}: $0.1 / 0.1 / 0.3$.
Cluster-scale vectors are adjusted (not necessarily monotonically) across the three settings for each clustering task.

\paragraph*{S4 Text.}
\label{S4_Text}
\textbf{Evaluation metrics and selection rules.}
We evaluated both CARVE-derived selectors and CVIs by treating each as a rule that selects $\widehat{k} \in \{3, \dots, 7\}$.
As the primary evaluation metric we take the adjusted Rand index (ARI) \cite{hubert1985comparing} between true labels and the clustering obtained at the $\widehat{k}$ selected by the respective criterion.
For CARVE, labels at the selected $k$ are obtained by consensus clustering the consensus matrix at the selected configuration; for CVIs, labels come from the base estimator at selected $k$.
The oracle reference $\mathrm{ARI}(k^\star)$ is the ARI obtained by running the same base estimator at the true $k^\star$. We additionally report the proportion of runs where the respective metric selected the correct $\widehat{k}$, i.e., where $\widehat{k}=k^\star$.

\textbf{CARVE selectors.}
We evaluated 6 ARI-based criteria: (i) \emph{stability} (agreement across subsample overlap), and (ii) \emph{generalizability} (agreement between labels predicted on held-out samples and labels on held-out samples). Each is combined with three selection rules:
\begin{itemize}
    \item \emph{Max rule} (\texttt{\_max}): select the $k$ maximizing the score.
    \item \emph{1SE rule} (\texttt{\_1se}): select the largest $k$ whose score is within one standard error of the maximum \cite{hastie2009elements}.
    \item \emph{Quantile rule} (\texttt{\_quant}): select the largest $k$ whose score lies within the 95\% confidence set of the maximum.
\end{itemize}
We additionally evaluated a consensus-based stability criterion based on the Gini index over the consensus matrix rows and an accuracy-based generalizability criterion, yielding 8 CARVE metrics in total.

\textbf{CVIs.}
We benchmark the Silhouette statistic \cite{rousseeuw1987silhouettes}, Davies--Bouldin \cite{davies2009cluster} (reported as $1/(1+\mathrm{DB})$ such that higher is better), Calinski--Harabasz \cite{calinski1974dendrite}, and the Gap statistic \cite{tibshirani2001estimating} (with 10 null-reference datasets).
For each, $\widehat{k} = \arg\max_{k \in \{3, \dots, 7\}}$ of the respective metric.

\textbf{Summary tables.}
Each per-task summary table (\nameref{S2_Table}--\nameref{S7_Table}) reports $B_{\text{datasets}}=20$ seeds at each of the three SNR settings (easy, medium, hard) for $k^\star = 5$. Within each clustering task we provide a paired sub-table giving mean $\mathrm{ARI}(\widehat{k})$ and the proportion of seeds for which $\widehat{k}=k^\star$ ($k$-recovery).
Rows are ranked by $\mathrm{ARI}(\widehat{k})$ mean; \textbf{bold} = best, \underline{underline} = second best (per column, excluding oracle).

\paragraph*{S1 Fig.}
Representative 2D PCA projections of each cluster shape at a hard SNR setting, with points colored by true cluster labels.

\label{S1_Fig}
\begin{figure}[H]
    \centering
    \includegraphics[width=1\linewidth]{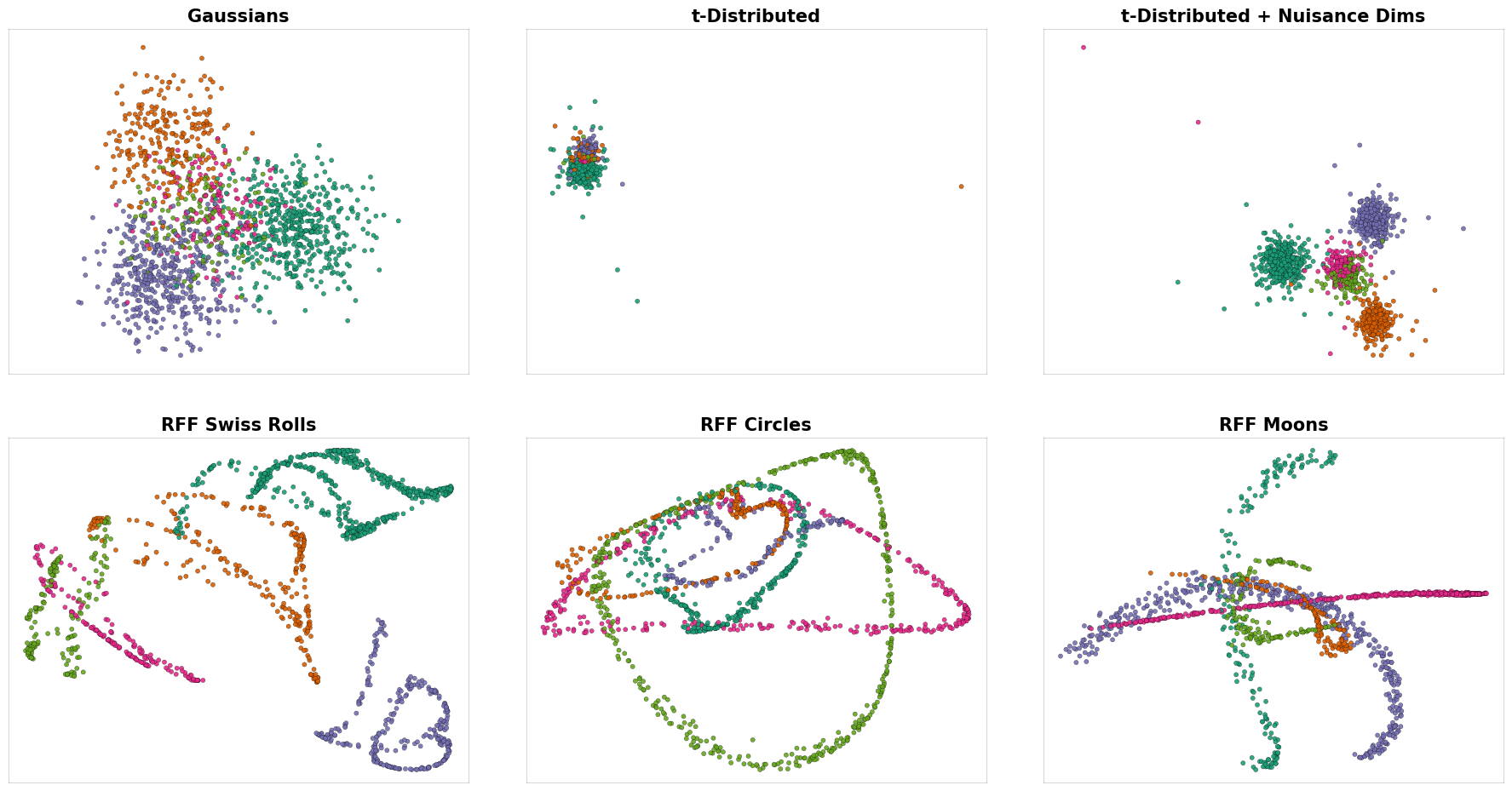}
    \caption{\textbf{Representative simulated datasets at $k^\star=5$.}
    Top row, left to right: Gaussian mixtures, correlated $t$-distributed mixtures, $t$-distributed mixtures with high-dimensional nuisance features. Bottom row: RFF-embedded swiss rolls, circles, and moons.}
    \label{fig:benchmarking_examples}
\end{figure}

\paragraph*{S2 Table.}
\label{S2_Table}
\textbf{Gaussian mixtures.} Mean $\mathrm{ARI}(\widehat{k})$ (top sub-table) and proportion of seeds where $\widehat{k} = k^\star$ (bottom sub-table) at easy, medium, and hard SNR settings for $k^\star = 5$.

\begin{table}[H]
\centering
\scriptsize
\setlength{\tabcolsep}{3pt}
\resizebox{\textwidth}{!}{%
\begin{tabular}{lccc}
\toprule
Metric & easy & medium & hard \\
\midrule
Baseline (Oracle) & 0.914 & 0.868 & 0.724 \\
\midrule
CARVE Stability (1SE) & \textbf{0.932} & 0.825 & 0.768 \\
CARVE Generalizability (1SE) & 0.868 & 0.725 & 0.719 \\
\midrule
Davies–Bouldin & 0.928 & \textbf{0.868} & \textbf{0.820} \\
Silhouette & 0.928 & \underline{0.858} & 0.785 \\
Gap Statistic & 0.851 & 0.737 & 0.507 \\
Calinski–Harabasz & 0.640 & 0.643 & 0.708 \\
\midrule
ARI (stab, quantile) & \textbf{0.932} & 0.852 & \underline{0.814} \\
Gini (stab) & 0.927 & 0.837 & 0.783 \\
ARI (stab, max) & \textbf{0.932} & 0.822 & 0.766 \\
ARI (gen, quantile) & 0.928 & 0.827 & 0.759 \\
ARI (gen, max) & 0.861 & 0.725 & 0.719 \\
Accuracy (gen) & 0.831 & 0.712 & 0.719 \\
\bottomrule
\end{tabular}%
\quad
\begin{tabular}{lccc}
\toprule
Metric & easy & medium & hard \\
\midrule
Baseline (Oracle) &  &  &  \\
\midrule
CARVE Stability (1SE) & \textbf{1.000} & 0.700 & 0.300 \\
CARVE Generalizability (1SE) & 0.550 & 0.100 & 0.000 \\
\midrule
Davies–Bouldin & 0.950 & \textbf{1.000} & \textbf{0.700} \\
Silhouette & 0.950 & \underline{0.900} & \underline{0.400} \\
Gap Statistic & 0.650 & 0.450 & 0.150 \\
Calinski–Harabasz & 0.000 & 0.000 & 0.000 \\
\midrule
ARI (stab, quantile) & \textbf{1.000} & \underline{0.900} & \underline{0.400} \\
Gini (stab) & 0.950 & 0.800 & \underline{0.400} \\
ARI (stab, max) & \textbf{1.000} & 0.650 & 0.250 \\
ARI (gen, quantile) & 0.950 & 0.600 & 0.150 \\
ARI (gen, max) & 0.550 & 0.100 & 0.000 \\
Accuracy (gen) & 0.450 & 0.050 & 0.000 \\
\bottomrule
\end{tabular}
}
\end{table}

\paragraph*{S3 Table.}
\label{S3_Table}
\textbf{$t$-distributed mixtures.} Mean $\mathrm{ARI}(\widehat{k})$ and $k$-recovery at easy/medium/hard for $k^\star = 5$.

\begin{table}[H]
\centering
\scriptsize
\setlength{\tabcolsep}{3pt}
\resizebox{\textwidth}{!}{%
\begin{tabular}{lccc}
\toprule
Metric & easy & medium & hard \\
\midrule
Baseline (Oracle) & 0.983 & 0.888 & 0.724 \\
\midrule
CARVE Stability (1SE) & 0.977 & 0.916 & \underline{0.695} \\
CARVE Generalizability (1SE) & 0.987 & 0.912 & 0.663 \\
\midrule
Gap Statistic & 0.985 & \textbf{0.943} & \textbf{0.805} \\
Silhouette & 0.985 & 0.860 & 0.584 \\
Davies–Bouldin & 0.972 & 0.875 & 0.544 \\
Calinski–Harabasz & 0.844 & 0.683 & 0.523 \\
\midrule
ARI (stab, quantile) & \underline{0.990} & \underline{0.924} & 0.687 \\
ARI (gen, quantile) & \underline{0.990} & \underline{0.924} & 0.687 \\
Gini (stab) & \textbf{0.991} & 0.916 & 0.693 \\
ARI (gen, max) & 0.987 & 0.912 & 0.663 \\
ARI (stab, max) & 0.948 & 0.916 & 0.659 \\
Accuracy (gen) & 0.912 & 0.879 & 0.603 \\
\bottomrule
\end{tabular}%
\quad
\begin{tabular}{lccc}
\toprule
Metric & easy & medium & hard \\
\midrule
Baseline (Oracle) &  &  &  \\
\midrule
CARVE Stability (1SE) & 0.450 & \underline{0.100} & \textbf{0.150} \\
CARVE Generalizability (1SE) & 0.550 & 0.050 & 0.100 \\
\midrule
Gap Statistic & 0.000 & 0.000 & 0.000 \\
Silhouette & 0.000 & \textbf{0.150} & \textbf{0.150} \\
Davies–Bouldin & 0.000 & \underline{0.100} & 0.050 \\
Calinski–Harabasz & 0.300 & 0.000 & 0.000 \\
\midrule
ARI (stab, quantile) & 0.000 & 0.000 & \textbf{0.150} \\
ARI (gen, quantile) & 0.000 & 0.000 & \textbf{0.150} \\
Gini (stab) & \textbf{0.650} & \underline{0.100} & \textbf{0.150} \\
ARI (gen, max) & \textbf{0.650} & 0.050 & 0.050 \\
ARI (stab, max) & 0.500 & \underline{0.100} & \textbf{0.150} \\
Accuracy (gen) & 0.500 & \underline{0.100} & \textbf{0.150} \\
\bottomrule
\end{tabular}
}
\end{table}

\paragraph*{S4 Table.}
\label{S4_Table}
\textbf{$t$-distributed mixtures with high-dimensional nuisance features.} Mean $\mathrm{ARI}(\widehat{k})$ and $k$-recovery at easy/medium/hard for $k^\star = 5$.

\begin{table}[H]
\centering
\scriptsize
\setlength{\tabcolsep}{3pt}
\resizebox{\textwidth}{!}{%
\begin{tabular}{lccc}
\toprule
Metric & easy & medium & hard \\
\midrule
Baseline (Oracle) & 0.981 & 0.848 & 0.721 \\
\midrule
CARVE Generalizability (1SE) & 0.967 & 0.921 & 0.850 \\
CARVE Stability (1SE) & 0.925 & 0.921 & \textbf{0.878} \\
\midrule
Davies–Bouldin & 0.982 & 0.849 & 0.717 \\
Gap Statistic & 0.981 & 0.831 & 0.695 \\
Silhouette & 0.981 & 0.838 & 0.681 \\
Calinski–Harabasz & 0.698 & 0.676 & 0.610 \\
\midrule
ARI (stab, quantile) & \textbf{0.994} & \underline{0.946} & 0.875 \\
ARI (gen, quantile) & \textbf{0.994} & \underline{0.946} & 0.875 \\
Gini (stab) & 0.974 & \textbf{0.947} & \underline{0.877} \\
ARI (gen, max) & 0.946 & 0.921 & 0.847 \\
ARI (stab, max) & 0.883 & 0.921 & 0.875 \\
Accuracy (gen) & 0.901 & 0.865 & 0.749 \\
\bottomrule
\end{tabular}%
\quad
\begin{tabular}{lccc}
\toprule
Metric & easy & medium & hard \\
\midrule
Baseline (Oracle) &  &  &  \\
\midrule
CARVE Generalizability (1SE) & 0.900 & \textbf{0.900} & \underline{0.400} \\
CARVE Stability (1SE) & 0.750 & \textbf{0.900} & 0.250 \\
\midrule
Davies–Bouldin & 0.750 & 0.550 & 0.050 \\
Gap Statistic & 0.750 & 0.550 & 0.000 \\
Silhouette & 0.850 & 0.700 & 0.100 \\
Calinski–Harabasz & 0.000 & 0.000 & 0.000 \\
\midrule
ARI (stab, quantile) & \textbf{1.000} & 0.650 & 0.100 \\
ARI (gen, quantile) & \textbf{1.000} & 0.700 & 0.100 \\
Gini (stab) & 0.950 & \textbf{0.900} & 0.200 \\
ARI (gen, max) & 0.800 & \textbf{0.900} & \textbf{0.500} \\
ARI (stab, max) & 0.550 & \textbf{0.900} & 0.300 \\
Accuracy (gen) & 0.550 & 0.450 & 0.100 \\
\bottomrule
\end{tabular}
}
\end{table}

\paragraph*{S5 Table.}
\label{S5_Table}
\textbf{RFF-embedded circles.} Mean $\mathrm{ARI}(\widehat{k})$ and $k$-recovery at easy/medium/hard for $k^\star = 5$.

\begin{table}[H]
\centering
\scriptsize
\setlength{\tabcolsep}{3pt}
\resizebox{\textwidth}{!}{%
\begin{tabular}{lccc}
\toprule
Metric & easy & medium & hard \\
\midrule
Baseline (Oracle) & 0.875 & 0.822 & 0.781 \\
\midrule
CARVE Stability (1SE) & \textbf{1.000} & \textbf{1.000} & \textbf{0.973} \\
CARVE Generalizability (1SE) & 0.984 & 0.942 & 0.781 \\
\midrule
Calinski–Harabasz & 0.721 & 0.729 & 0.652 \\
Silhouette & 0.722 & 0.725 & 0.651 \\
Gap Statistic & 0.722 & 0.722 & 0.651 \\
Davies–Bouldin & 0.722 & 0.722 & 0.651 \\
\midrule
ARI (stab, quantile) & \textbf{1.000} & \textbf{1.000} & 0.947 \\
ARI (stab, max) & 0.988 & 0.971 & \underline{0.951} \\
Gini (stab) & 0.988 & 0.949 & 0.941 \\
Accuracy (gen) & 0.984 & 0.942 & 0.802 \\
ARI (gen, max) & 0.984 & 0.942 & 0.791 \\
ARI (gen, quantile) & 0.964 & 0.914 & 0.728 \\
\bottomrule
\end{tabular}%
\quad
\begin{tabular}{lccc}
\toprule
Metric & easy & medium & hard \\
\midrule
Baseline (Oracle) &  &  &  \\
\midrule
CARVE Stability (1SE) & \textbf{1.000} & \textbf{1.000} & \textbf{0.700} \\
CARVE Generalizability (1SE) & 0.900 & 0.600 & 0.100 \\
\midrule
Calinski–Harabasz & 0.050 & 0.100 & 0.100 \\
Silhouette & 0.000 & 0.000 & 0.000 \\
Gap Statistic & 0.000 & 0.000 & 0.000 \\
Davies–Bouldin & 0.000 & 0.000 & 0.000 \\
\midrule
ARI (stab, quantile) & \textbf{1.000} & \textbf{1.000} & \underline{0.650} \\
ARI (stab, max) & 0.900 & 0.850 & 0.600 \\
Gini (stab) & 0.900 & 0.750 & 0.550 \\
Accuracy (gen) & 0.900 & 0.600 & 0.050 \\
ARI (gen, max) & 0.900 & 0.600 & 0.100 \\
ARI (gen, quantile) & 0.850 & 0.600 & 0.050 \\
\bottomrule
\end{tabular}
}
\end{table}

\paragraph*{S6 Table.}
\label{S6_Table}
\textbf{RFF-embedded moons.} Mean $\mathrm{ARI}(\widehat{k})$ and $k$-recovery at easy/medium/hard for $k^\star = 5$.

\begin{table}[H]
\centering
\scriptsize
\setlength{\tabcolsep}{3pt}
\resizebox{\textwidth}{!}{%
\begin{tabular}{lccc}
\toprule
Metric & easy & medium & hard \\
\midrule
Baseline (Oracle) & 0.891 & 0.861 & 0.802 \\
\midrule
CARVE Stability (1SE) & \textbf{1.000} & \textbf{0.994} & \textbf{1.000} \\
CARVE Generalizability (1SE) & 0.968 & 0.874 & 0.940 \\
\midrule
Silhouette & 0.719 & 0.740 & 0.570 \\
Davies–Bouldin & 0.719 & 0.742 & 0.551 \\
Gap Statistic & 0.719 & 0.742 & 0.541 \\
Calinski–Harabasz & 0.719 & 0.730 & 0.538 \\
\midrule
ARI (stab, quantile) & \textbf{1.000} & 0.966 & \textbf{1.000} \\
ARI (stab, max) & 0.997 & \underline{0.983} & 0.985 \\
Gini (stab) & 0.989 & 0.978 & 0.985 \\
ARI (gen, max) & 0.965 & 0.858 & 0.945 \\
Accuracy (gen) & 0.958 & 0.850 & 0.849 \\
ARI (gen, quantile) & 0.912 & 0.796 & 0.727 \\
\bottomrule
\end{tabular}%
\quad
\begin{tabular}{lccc}
\toprule
Metric & easy & medium & hard \\
\midrule
Baseline (Oracle) &  &  &  \\
\midrule
CARVE Stability (1SE) & \textbf{1.000} & \textbf{0.900} & \textbf{1.000} \\
CARVE Generalizability (1SE) & 0.750 & 0.350 & 0.400 \\
\midrule
Silhouette & 0.000 & 0.000 & 0.000 \\
Davies–Bouldin & 0.000 & 0.000 & 0.000 \\
Gap Statistic & 0.000 & 0.000 & 0.000 \\
Calinski–Harabasz & 0.050 & 0.100 & 0.000 \\
\midrule
ARI (stab, quantile) & \textbf{1.000} & \underline{0.850} & \textbf{1.000} \\
ARI (stab, max) & 0.950 & \underline{0.850} & 0.900 \\
Gini (stab) & 0.900 & 0.800 & 0.900 \\
ARI (gen, max) & 0.700 & 0.300 & 0.400 \\
Accuracy (gen) & 0.650 & 0.200 & 0.300 \\
ARI (gen, quantile) & 0.600 & 0.150 & 0.300 \\
\bottomrule
\end{tabular}
}
\end{table}

\paragraph*{S7 Table.}
\label{S7_Table}
\textbf{RFF-embedded Swiss rolls.} Mean $\mathrm{ARI}(\widehat{k})$ and $k$-recovery at easy/medium/hard for $k^\star = 5$.

\begin{table}[H]
\centering
\scriptsize
\setlength{\tabcolsep}{3pt}
\resizebox{\textwidth}{!}{%
\begin{tabular}{lccc}
\toprule
 & \multicolumn{3}{c}{$k^* = 5$} \\
\cmidrule(lr){2-4}
Metric & easy & medium & hard \\
\midrule
Baseline (Oracle) & 0.982 & 0.860 & 0.752 \\
\midrule
CARVE Stability (1SE) & \textbf{0.967} & \underline{0.805} & \textbf{0.788} \\
CARVE Generalizability (1SE) & \underline{0.962} & 0.793 & \underline{0.786} \\
\midrule
Silhouette & 0.943 & \textbf{0.852} & 0.766 \\
Calinski–Harabasz & 0.853 & 0.755 & 0.734 \\
Davies–Bouldin & 0.804 & 0.741 & 0.700 \\
Gap Statistic & 0.738 & 0.741 & 0.703 \\
\midrule
ARI (stab, quantile) & 0.952 & 0.778 & \underline{0.786} \\
ARI (gen, quantile) & 0.947 & 0.779 & 0.778 \\
ARI (gen, max) & 0.932 & 0.793 & 0.778 \\
Accuracy (gen) & 0.932 & 0.784 & 0.760 \\
ARI (stab, max) & 0.890 & 0.775 & 0.772 \\
Gini (stab) & 0.876 & 0.763 & 0.780 \\
\bottomrule
\end{tabular}
\begin{tabular}{lccc}
\toprule
 & \multicolumn{3}{c}{$k^* = 5$} \\
\cmidrule(lr){2-4}
Metric & easy & medium & hard \\
\midrule
Baseline (Oracle) &  &  &  \\
\midrule
CARVE Stability (1SE) & \textbf{0.800} & \textbf{0.150} & \underline{0.100} \\
CARVE Generalizability (1SE) & \textbf{0.800} & \textbf{0.150} & \underline{0.100} \\
\midrule
Silhouette & 0.700 & \textbf{0.150} & 0.050 \\
Calinski–Harabasz & 0.450 & 0.000 & \textbf{0.150} \\
Davies–Bouldin & 0.150 & 0.000 & 0.000 \\
Gap Statistic & 0.000 & 0.000 & 0.000 \\
\midrule
ARI (stab, quantile) & 0.750 & 0.050 & \underline{0.100} \\
ARI (gen, quantile) & 0.750 & 0.050 & 0.050 \\
ARI (gen, max) & 0.550 & \textbf{0.150} & \underline{0.100} \\
Accuracy (gen) & 0.550 & 0.100 & \underline{0.100} \\
ARI (stab, max) & 0.450 & 0.100 & \underline{0.100} \\
Gini (stab) & 0.400 & 0.100 & \underline{0.100} \\
\bottomrule
\end{tabular}
}
\end{table}


\paragraph*{S8 Table.}
\label{S8_Table}
\textbf{Scaling --- Gaussian mixtures, sample-size axis: $\mathrm{ARI}(\widehat{k})$ and proportion where $\widehat{k} = k^\star$.}
Mean ARI of the selected $k$ (left) and proportion of runs with $\widehat{k}=k^\star$ (right), evaluated at three sample sizes for Gaussian mixtures (KMeans estimator). $B=20$ seeds per $(n, k^\star)$ cell. \textbf{bold} = best, \underline{underline} = second best per column (excluding oracle).

\begin{table}[H]
\centering
\scriptsize
\setlength{\tabcolsep}{3pt}
\resizebox{\textwidth}{!}{%
\begin{tabular}{lccc}
\toprule
Metric & 1000 & 5500 & 10000 \\
\midrule
Baseline (Oracle) & 0.792 & 0.752 & 0.726 \\
\midrule
CARVE Stability (1SE) & 0.737 & 0.767 & 0.733 \\
CARVE Generalizability (1SE) & 0.682 & 0.714 & 0.667 \\
\midrule
Silhouette & \textbf{0.821} & \textbf{0.802} & \textbf{0.762} \\
Davies–Bouldin & \underline{0.806} & \underline{0.788} & 0.735 \\
Calinski–Harabasz & 0.618 & 0.597 & 0.544 \\
Gap Statistic & 0.552 & 0.562 & 0.533 \\
\midrule
ARI (stab, quantile) & 0.745 & 0.785 & \underline{0.752} \\
Gini (stab) & 0.734 & 0.767 & 0.741 \\
ARI (stab, max) & 0.737 & 0.767 & 0.723 \\
ARI (gen, quantile) & 0.729 & 0.743 & 0.708 \\
ARI (gen, max) & 0.682 & 0.714 & 0.664 \\
Accuracy (gen) & 0.653 & 0.673 & 0.571 \\
\bottomrule
\end{tabular}%
\quad
\begin{tabular}{lccc}
\toprule
Metric & 1000 & 5500 & 10000 \\
\midrule
Baseline (Oracle) &  &  &  \\
\midrule
CARVE Stability (1SE) & 0.250 & 0.550 & 0.600 \\
CARVE Generalizability (1SE) & 0.150 & 0.300 & 0.350 \\
\midrule
Silhouette & \underline{0.750} & \textbf{0.750} & \textbf{0.800} \\
Davies–Bouldin & \textbf{0.850} & \textbf{0.750} & \textbf{0.800} \\
Calinski–Harabasz & 0.000 & 0.000 & 0.000 \\
Gap Statistic & 0.000 & 0.000 & 0.000 \\
\midrule
ARI (stab, quantile) & 0.400 & 0.650 & \textbf{0.800} \\
Gini (stab) & 0.250 & 0.550 & 0.700 \\
ARI (stab, max) & 0.250 & 0.550 & 0.550 \\
ARI (gen, quantile) & 0.300 & 0.550 & 0.550 \\
ARI (gen, max) & 0.150 & 0.300 & 0.300 \\
Accuracy (gen) & 0.150 & 0.100 & 0.000 \\
\bottomrule
\end{tabular}%
}
\end{table}

\paragraph*{S9 Table.}
\label{S9_Table}
\textbf{Scaling --- Gaussian mixtures, dimensionality axis: $\mathrm{ARI}(\widehat{k})$ and proportion where $\widehat{k} = k^\star$.}
As S8, but varying the feature dimension $p$ for Gaussian mixtures (KMeans).

\begin{table}[H]
\centering
\scriptsize
\setlength{\tabcolsep}{3pt}
\resizebox{\textwidth}{!}{%
\begin{tabular}{lccc}
\toprule
Metric & 50 & 525 & 1000 \\
\midrule
Baseline (Oracle) & 0.800 & 0.747 & 0.819 \\
\midrule
CARVE Stability (1SE) & 0.759 & 0.693 & 0.737 \\
CARVE Generalizability (1SE) & 0.691 & 0.309 & 0.310 \\
\midrule
Silhouette & \textbf{0.810} & \textbf{0.847} & \textbf{0.909} \\
Davies–Bouldin & \underline{0.792} & \underline{0.780} & \underline{0.908} \\
Calinski–Harabasz & 0.596 & 0.660 & 0.708 \\
Gap Statistic & 0.545 & 0.606 & 0.666 \\
\midrule
ARI (stab, quantile) & 0.772 & 0.731 & 0.773 \\
Gini (stab) & 0.755 & 0.720 & 0.737 \\
ARI (stab, max) & 0.758 & 0.684 & 0.737 \\
ARI (gen, quantile) & 0.718 & 0.394 & 0.278 \\
ARI (gen, max) & 0.684 & 0.309 & 0.293 \\
Accuracy (gen) & 0.673 & 0.297 & 0.174 \\
\bottomrule
\end{tabular}%
\quad
\begin{tabular}{lccc}
\toprule
Metric & 50 & 525 & 1000 \\
\midrule
Baseline (Oracle) &  &  &  \\
\midrule
CARVE Stability (1SE) & 0.400 & 0.000 & 0.000 \\
CARVE Generalizability (1SE) & 0.250 & 0.000 & 0.000 \\
\midrule
Silhouette & \underline{0.750} & \textbf{0.550} & \textbf{0.500} \\
Davies–Bouldin & \textbf{0.850} & \textbf{0.550} & \textbf{0.500} \\
Calinski–Harabasz & 0.000 & 0.000 & 0.000 \\
Gap Statistic & 0.000 & 0.000 & 0.000 \\
\midrule
ARI (stab, quantile) & 0.350 & 0.200 & 0.200 \\
Gini (stab) & 0.400 & 0.000 & 0.000 \\
ARI (stab, max) & 0.400 & 0.000 & 0.000 \\
ARI (gen, quantile) & 0.350 & 0.300 & 0.150 \\
ARI (gen, max) & 0.150 & 0.000 & 0.000 \\
Accuracy (gen) & 0.150 & 0.000 & 0.000 \\
\bottomrule
\end{tabular}%
}
\end{table}

\paragraph*{S10 Table.}
\label{S10_Table}
\textbf{Scaling --- CARVE runtime (seconds) across the Gaussian-mixture scaling experiments.}
Mean (sd) seconds for one \texttt{CARVE.fit()} call for stability and generalizability, aggregated across $B=20$ seeds at three axis values per experiment. Stability mode is consistently the cheaper of the two. However, runtimes heavily depend on clustering method and classifier used.

\begin{table}[H]
\centering
\scriptsize
\setlength{\tabcolsep}{3pt}
\resizebox{\textwidth}{!}{
\begin{tabular}{llccc}
\toprule
Setting & Mode & low & mid & high \\
\midrule
Gaussian (samples) (n\_total = 1000 / 5500 / 10000) & CARVE Stability & 2.18 (0.09) & 14.65 (1.17) & 42.90 (2.98) \\
 & CARVE Generalizability & 15.44 (0.24) & 86.14 (1.56) & 179.24 (7.78) \\
\midrule
Gaussian (dimensionality) (p = 50 / 525 / 1000) & CARVE Stability & 2.73 (0.19) & 27.47 (2.44) & 72.51 (5.38) \\
 & CARVE Generalizability & 20.06 (0.50) & 62.44 (3.73) & 104.87 (2.46) \\
\bottomrule
\end{tabular}
}
\end{table}

\paragraph*{S5 Text.}
\label{S5_Text}
\textbf{Scaling experiments: experimental setup and detailed results.}

\textbf{Experimental design.}
We evaluated how $\mathrm{ARI}(\widehat{k})$ and runtime scale along two axes for isotropic Gaussian mixtures clustered with KMeans. Sample size $n$ and feature dimension $p$ were varied over three values each: $n \in \{1{,}000;\, 5{,}500;\, 10{,}000\}$ and $p \in \{50;\, 525;\, 1{,}000\}$. For each setting we generated $B=20$ datasets at $k^\star = 5$ and ran the same suite of CARVE selectors and CVIs as in the synthetic benchmarks (\nameref{S4_Text}).

\textbf{Accuracy.}
Stability-based selection rules remained essentially unchanged across both axes (\nameref{S2_Fig}). Generalizability-based selection rules, however, degraded at medium-to-large $p$ (see \nameref{S9_Table}). The cause is that the default classifier used to score generalizability is a random forest with 100 trees; in high dimensions this ensemble is no longer able to capture the geometry of the embedded clusters from a single subsample. We thus recommend that the number of trees should be increased as $p$ grows, or a different classifier should be supplied via the \texttt{classifier} argument of \texttt{CARVE()}.

\textbf{Runtime.}
Runtime is dominated by the cost of the underlying clustering algorithm and the classifier; the resampling loop itself adds modest overhead (\nameref{S3_Fig}, \nameref{S10_Table}). Generalizability mode is consistently slower than stability mode because it additionally fits a classifier on each subsample and evaluates it on the held-out complement.

\textbf{Per-experiment tables.}
Mean $\mathrm{ARI}(\widehat{k})$ and $k$-recovery are tabulated per setting in \nameref{S8_Table} (sample-size axis) and \nameref{S9_Table} (dimensionality axis); runtime is given in \nameref{S10_Table}.

\paragraph*{S2 Fig.}
\label{S2_Fig}
\textbf{Scaling: $\mathrm{ARI}(\widehat{k})$ accuracy as a function of $n$ and $p$ for Gaussian mixtures at $k^\star = 5$.}
Mean $\mathrm{ARI}(\widehat{k})$ for KMeans on Gaussian mixtures across $B=20$ seeds. Left: vs.\ sample size $n$; right: vs.\ feature dimension $p$. Curves show CARVE stability (1-SE) and generalizability (1-SE) against the oracle-$k^\star$ baseline. Runtime curves are in \nameref{S3_Fig}, and tabulated per-setting values are in \nameref{S8_Table}--\nameref{S10_Table}.

\begin{figure}[h!]
    \centering
    \includegraphics[width=1\linewidth]{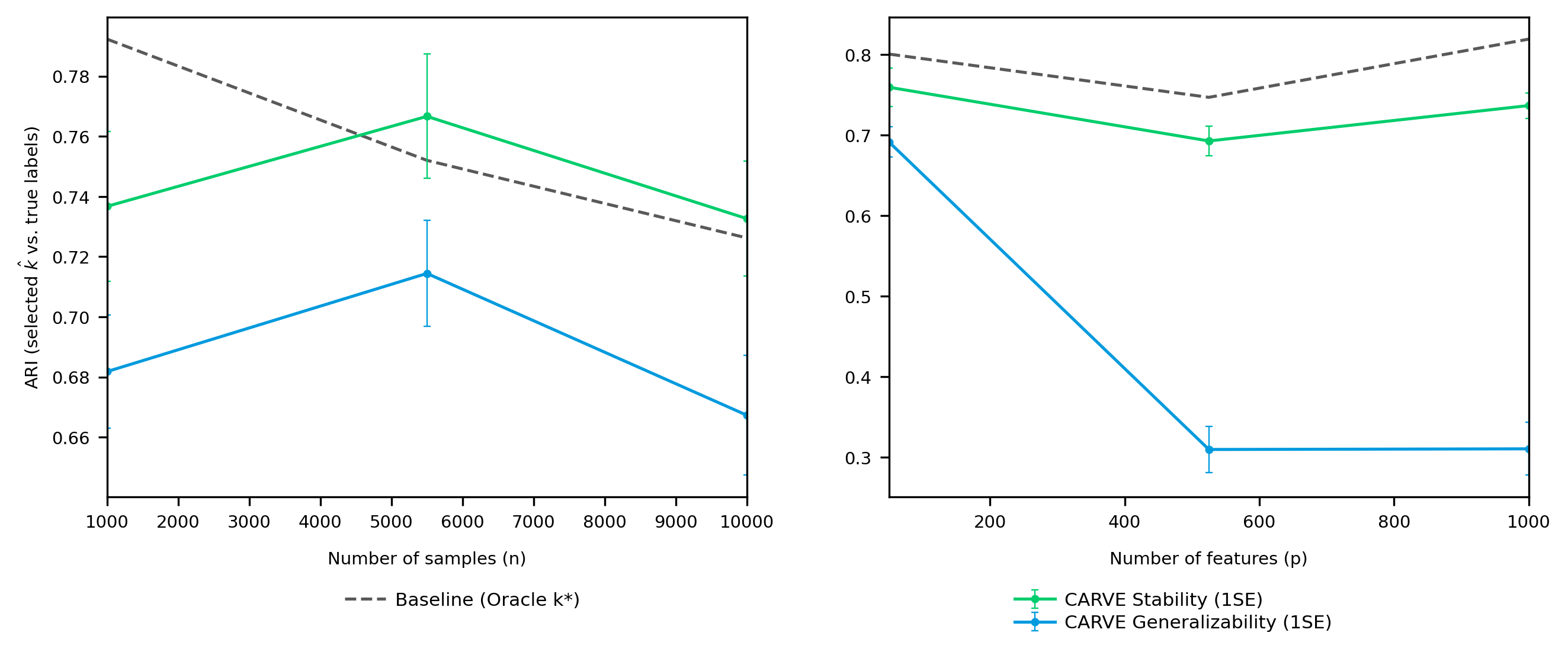}
    \label{fig:scaling_main_k5}
\end{figure}

\paragraph*{S3 Fig.}
\label{S3_Fig}
\textbf{CARVE runtime on Gaussian mixtures as a function of $n$ and $p$ at $k^\star = 5$.}
Runtime in seconds (log scale) for one \texttt{CARVE.fit()} call for stability and generalizability, aggregated across $B=20$ seeds. Left: vs.\ sample size $n$. Right: vs.\ feature dimension $p$. Tabular form is given in \nameref{S10_Table}.
\begin{figure}[h!]
    \centering
    \includegraphics[width=1\linewidth]{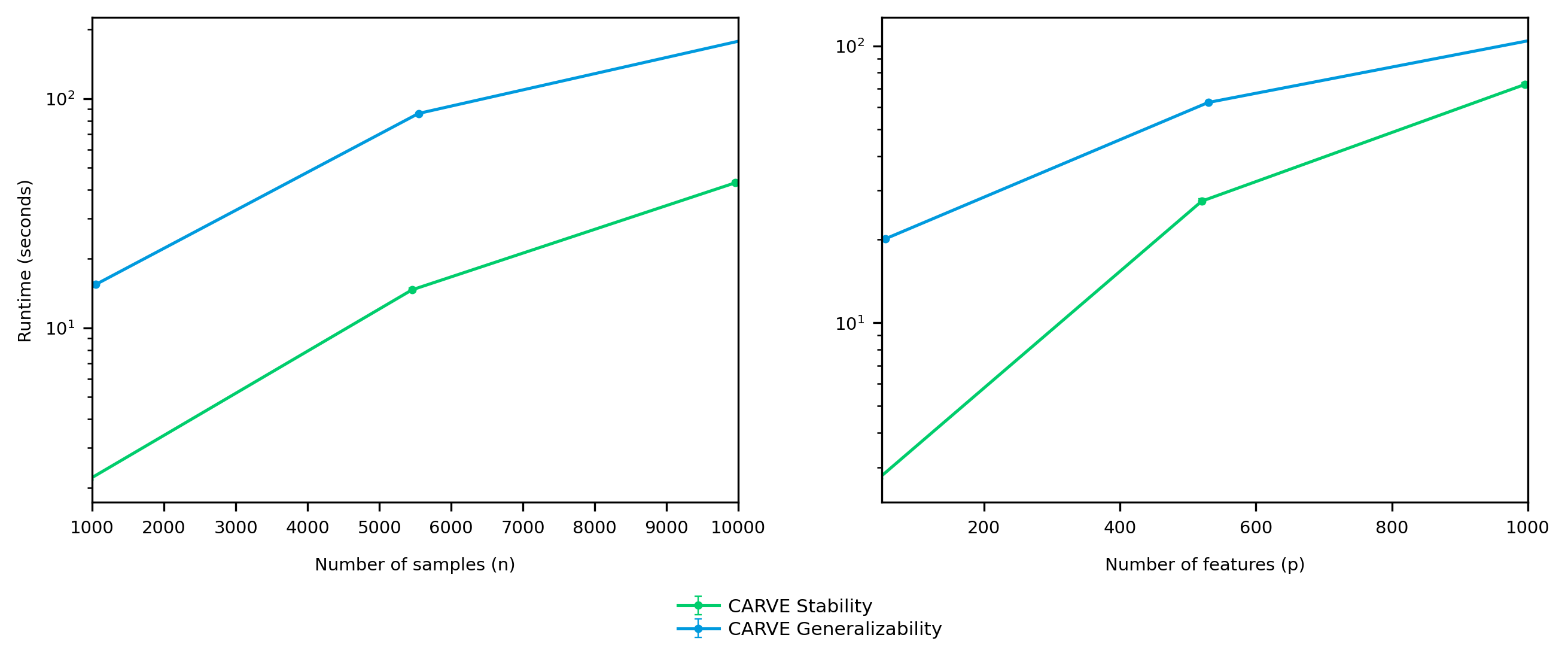}
    \label{fig:scaling_runtime_k5}
\end{figure}

\paragraph*{S6 Text.}
\label{S6_Text}
\textbf{Case study 1: Klein droplet-based scRNA-seq data.}

\textbf{Data source.}
Raw gene-count matrices for the four experimental conditions in \cite{klein2015droplet} were obtained from GEO accessions GSM1599494 (day 0, +LIF control), GSM1599497 (day 2 post-LIF withdrawal), GSM1599498 (day 4), and GSM1599499 (day 7), each provided as a bz2-compressed CSV with genes as rows and cells as columns.

\textbf{Alignment and cleaning.}
The four per-condition matrices were outer-joined on their gene indices (missing entries filled with 0), non-numeric entries were coerced to 0, and duplicate gene identifiers were removed (retaining the first occurrence). This yielded a combined matrix of 24{,}175 genes by 2{,}717 cells.

\textbf{Normalization and transformation.}
Cells were normalized to a common library size of $10^{4}$ counts (\texttt{sc.pp.normalize\_total(target\_sum=1e4)}) and then log-transformed via $\log(1+x)$ (\texttt{sc.pp.log1p}).

\textbf{Highly variable gene selection.}
We selected 2{,}000 highly variable genes using \texttt{sc.pp.highly\_variable\_genes}, \texttt{flavor='seurat'}, \texttt{n\_top\_genes=2000}, \texttt{batch\_key='condition'}, with HVG ranking computed per condition to avoid confounding by developmental stage.

\textbf{Stratified subsampling.}
To keep case-study runtime tractable, we drew a stratified 50\% subsample using \texttt{StratifiedShuffleSplit} (\texttt{test\_size=0.5}, \texttt{random\_state=42}), stratified by condition so that the relative proportions of d0/d2/d4/d7 cells were preserved. The final analysis matrix consists of 1{,}358 cells (466 d0, 152 d2, 341 d4, 399 d7) by 2{,}000 genes; reference labels are the experimental-condition labels stored in \texttt{adata.obs['condition']}.

\textbf{Clustering algorithm grid and CVI comparison.}

For the Klein dataset we evaluated two clustering algorithms --- Ward agglomerative clustering and spectral clustering with self-tuning affinity \cite{zelnik2004self} --- each over $k \in \{2,\dots,10\}$, and computed four CVIs: Silhouette, Calinski--Harabasz (CH), a transformed Davies--Bouldin ($\mathrm{DB}' = 1/(1+\mathrm{DB})$) score, and the gap statistic (\textbf{Fig~\ref{fig:klein_results}(E)}). The Silhouette and CH indices are both maximized at $k=2$ under Ward agglomerative clustering; DB also selects $k=2$ under Ward agglomerative clustering. Inspecting the $k=2$ solution (\textbf{Fig~\ref{fig:klein_results}(C)}) confirms that the CVIs collapse four reported stages into two coarse groups.

By contrast, CARVE's generalizability ARI with the 1-SE rule selects Ward agglomerative clustering at $k=4$ and CARVE's cluster-level stability diagnostics (\textbf{Fig~\ref{fig:klein_results}(D)}) further distinguish the two end-state clusters (d0, d7; higher stability) from the two transitional clusters (d2, d4; lower, heterogeneous stability). 

\paragraph*{S7 Text.}
\label{S7_Text}
\textbf{Case study 2: Levine 32-dim mass cytometry data.}

\textbf{Data source.}
The Levine 32-dim benchmark from \cite{levine2015data} was obtained through the \texttt{HDCytoData} Bioconductor package (\texttt{HDCytoData::Levine\_32dim\_SE}) via \texttt{rpy2}. The full dataset comprises 265{,}627 cells measured on 39 protein markers, with manually gated population labels stored in \texttt{colData(sce)[['population\_id']]}.

\textbf{Marker and label filtering.}
We restricted the marker set to the 32 ``type'' markers (\texttt{marker\_class == 2}), which define cell identity rather than functional state. Cells with population ID~15 --- the largest manually gated population (161{,}443 cells), labeled as uncharacterized in the original benchmark --- were removed, along with any unlabeled cells. This yielded 104{,}184 cells across the 14 annotated populations.

\textbf{Transformation and scaling.}
Marker intensities were transformed with \texttt{arcsinh} at cofactor~5 and each marker was then scaled to its 10th--90th percentile range using \texttt{RobustScaler(quantile\_range=(10, 90))}. 

\textbf{Stratified subsampling.}
To keep runtime tractable we drew a stratified subsample of 5{,}000 cells with \texttt{StratifiedShuffleSplit} (\texttt{test\_size}$\approx$\texttt{0.048}, \texttt{random\_state=42}), stratified by \texttt{population\_id} so that minority populations were preserved in proportion. The final analysis matrix consists of 5{,}000 cells by 32 markers.

\textbf{Clustering algorithm grid and CVI comparison.}

For the Levine dataset we evaluated KMeans and spectral clustering with self-tuning affinity for $k \in \{7,\dots,17\}$ and computed four CVIs (\textbf{Fig~\ref{fig:levine_results}(E)}). Three of the four (Silhouette, Davies--Bouldin, Calinski--Harabasz) suggest KMeans at $k=7$ ($\mathrm{ARI}=0.62$ against the 14 reported populations); the gap statistic instead selects spectral clustering at $k=15$ ($\mathrm{ARI}=0.49$). 

CARVE's stability and generalizability 1-SE rules both select $k=10$ but agree on different base estimators --- spectral clustering for stability and KMeans for generalizability. This indicates that the two axes capture notions of good clustering (reproducibility under resampling versus out-of-sample predictability) which combined add up to a more holistic understanding of the clustering solutions. The selected $k=10$ is robust to the choice of metric, but we prefer the stability-based approach as this approach generally performed better during benchmarking. Full per-criterion ARI values are reported alongside \textbf{Fig~\ref{fig:levine_results}(F)}.

To interpret the partitions, we mapped the integer \texttt{population\_id} labels to cell-type names using the population-name table accompanying the benchmark \cite{weber2016comparison}. Cross-tabulating the reported labels against the CVI ($k=7$) and CARVE ($k=10$) partitions shows that the gain at $k=10$ stems from resolving the three major lymphocyte populations --- CD4$^+$ T, CD8$^+$ T, and CD16$^-$ NK cells --- which the CVI-selected clustering pools into a single cluster. The rarest progenitor and stem-cell subsets (e.g.\ CD34$^+$CD38$^{\mathrm{lo}}$ HSCs and plasma B cells) remain merged under both partitions.

\paragraph*{S4 Fig.}
\label{S4_Fig}
\textbf{Example CARVE output on the Levine 32-dim mass cytometry dataset \cite{levine2015data}.}
Visual output returned by a \texttt{CARVE.fit()} run that swept KMeans and self-tuning spectral clustering across $k \in \{7,\dots,17\}$. \textbf{(A)} Stability ARI as a function of $k$ for each estimator. The method selected by the respective selection-rule is marked by a vertical line. \textbf{(B)} Consensus matrix for a selected configuration: clear block-diagonal structure indicates samples that consistently co-cluster across resamples, while higher off-diagonal values indicate ambiguous samples and clusters. \textbf{(C)} Generalizability ARI as a function of $k$ for the same estimators, again with the 1-SE-selected model highlighted. \textbf{(D)} Per-cluster stability scores (violin plot) for the selected partition, exposing which clusters are stable across resamples and which are heterogeneous or dubious. \textbf{(E)} t-SNE \cite{vandermaaten2008visualizing} embedding colored by the CARVE-selected consensus labels. Dubious samples are larger with higher opacity. \textbf{(F)} t-SNE embedding marked by CARVE-selected consensus cluster-assignment. Highlighted samples are more spurious.

\begin{figure}[H]
    \centering
    \includegraphics[width=1\linewidth]{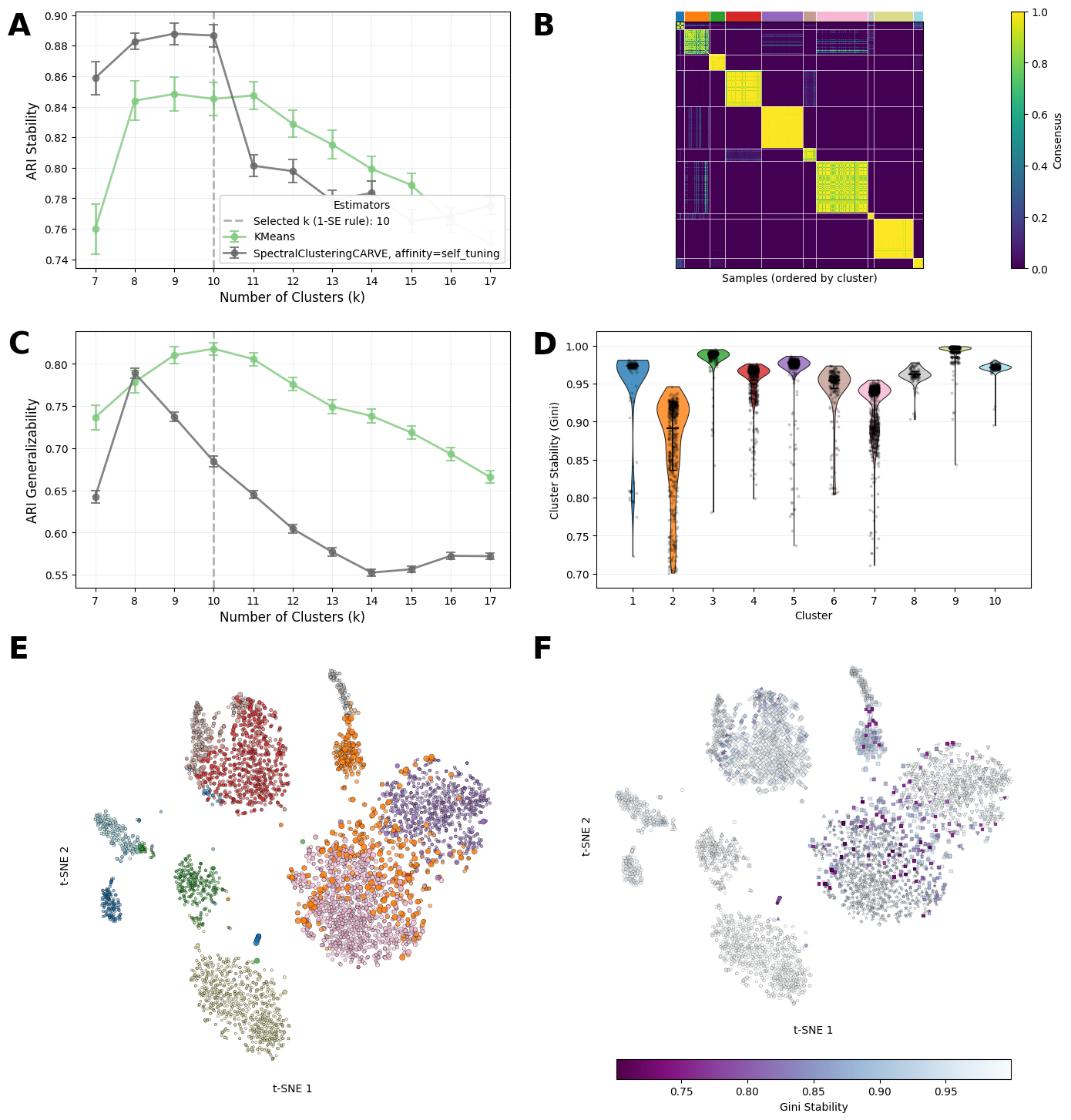}
    \label{fig:carve_output_levine}
\end{figure}

%
%

\newpage

\bibliography{plos_bibtex}

\end{document}